\documentclass[iop,onecolumn,numberedappendix]{emulateapj}
\usepackage[utf8]{inputenc}
\usepackage{color,ulem}
\usepackage{graphicx}
\usepackage{amssymb}
\usepackage{bm}
\usepackage{amsmath}
\usepackage[colorlinks=true,linkcolor=blue,citecolor=blue,%
urlcolor=blue,linktocpage=true]{hyperref}

\def\approxprop{%
  \def\p{%
    \setbox0=\vbox{\hbox{$\propto$}}%
    \ht0=0.6ex \box0 }%
  \def\s{%
    \vbox{\hbox{$\sim$}}%
  }%
  \mathrel{\raisebox{0.7ex}{%
      \mbox{$\underset{\s}{\p}$}%
    }}%
}

\date{September 2019}

\begin{document}
\title{Radioactive heating rate of r-process elements  and macronova light curve}
\author{Kenta Hotokezaka$^1$ and
Ehud Nakar$^2$}

\affil{
$^1$Department of Astrophysical Sciences, Princeton University, Princeton, 4 Ivy Ln, NJ 08544, USA\\
$^2$Raymond and Beverly Sackler School of Physics \& Astronomy, Tel Aviv University, Tel Aviv 69978, Israel\\
}

\begin{abstract}
We study the heating rate of r-process nuclei and thermalization of decay products in neutron star merger ejecta and macronova (kilonova) light curves.  Thermalization of charged decay products, i.e., electrons, $\alpha$-particles, and fission fragments is calculated according to their injection energy. The $\gamma$-ray thermalization processes are also properly calculated by taking the $\gamma$-ray spectrum of each decay into account. We show that the $\beta$-decay heating rate at later times approaches a power-law decline as $\propto t^{-2.8}$, which agrees with the result of \cite{waxman2019}. 
We present a new analytic model to calculate macronova light curves, in which the density structure of the ejecta is accounted for.   
 We demonstrate that the observed bolometric light curve and temperature evolution of the macronova associated with GW170817 are reproduced well by the $\beta$-decay heating rate with the solar r-process abundance pattern. We interpret the break in the observed bolometric light curve around a week as a result of the diffusion wave crossing a significant part of the ejecta rather than a thermalization break.
 We also show that the time-weighted integral of the bolometric light curve (Katz integral) is useful to provide an estimate of the total r-process mass from the observed data, which is independent of the highly uncertain radiative transfer. For the macronova in GW170817, the ejecta mass  is robustly estimated as $\approx 0.05M_{\odot}$ for $A_{\rm min}\leq 72$ and $85\leq A_{\rm min}\leq 130$ with the solar r-process abundance pattern. The code for computation of the heating rate and light curve for given initial nuclear abundances is publicly available.
\end{abstract}

\section{Introduction}
A neutron star merger ejects a considerable amount of neutron-rich material at subrelativistic velocities. The physical conditions of merger ejecta are 
ideal for  r-process nucleosynthesis \citep{Lattimer1974,lattimer1976ApJ,symbalisty1982ApL,freiburghaus1999ApJ} and it has been suggested as the origin of r-process elements of the solar system (see, e.g., \citealt{Thielemann2017,Hotokezaka2018IJMPD} and references therein).
Radioactive decay of r-process nuclei synthesized in  merger ejecta  produce long-term heat, which powers an uv/optical/IR transient, a so-called macronova (kilonova) \citep{li1998ApJ,Metzger2010}.  The first binary neutron star merger event, GW170817, was indeed accompanied by a macronova \citep{Abbott2017ApJ}. The light curves and spectra of this macronova are largely consistent with r-process-powered macronova models \citep{Andreoni2017,Arcavi2017Natur,Coulter2017,Cowperthwaite2017,Drout2017,Evans2017,Kasliwal2017,Pian2017,Smartt2017,tanvir2017ApJ,Utsumi2017}. 


 The radioactive power of $\beta$-decay of r-process nuclei declines with time with a characteristic power law \citep{Metzger2010,Goriely2011,Roberts2011ApJ,korobkin2012MNRAS,perego2014MNRAS,wanajo2014ApJ}. This behavior results from the existence of many different $\beta$-decay chains statistically contributing to the heat \citep{hotokezaka2017MNRAS}. The exact shape of the radioactive power, however, depends on the ejecta composition, which is primarily determined by the initial electron fraction.  \cite{lippuner2015ApJ} and \cite{wanajo2018} systematically studied the radioactive power under  various ejecta conditions. In addition to $\beta$-decay, $\alpha$-decay and spontaneous fission of heavy nuclei ($A\gtrsim 220$) are suggested to predominantly power macronovae
\citep{wanajo2014ApJ,Hotokezaka2016MNRAS,zhu2018,wu2019}. For instance,  $\alpha$-decaying nuclei with an atomic mass number of $A=222$-$225$ \citep{wu2019} and spontaneous fission of $^{254}$Cf may be an important heat source of  macronovae \citep{zhu2018,wu2019}.
Thermalization of decay products, $\gamma$-rays, electrons, $\alpha$-particles, and fission fragments,  also plays important roles for the heating rate \citep{Hotokezaka2016MNRAS,barnes2016,kasen2018,waxman2019}. 
Thermalization of charged particles is first considered by \cite{barnes2016}. More recently,  \cite{kasen2018} and \cite{waxman2019}  analytically showed that the decline of the late-time heating rate has a characteristic slope. However, there is a discrepancy between the two papers. \cite{kasen2018} obtain a slope of $\approx -2.3$ while  \cite{waxman2019} obtain a steeper slope of $\approx -2.8$. 

Deriving a macronova light curve of ejecta with known properties (mass, composition and velocity profiles) requires the calculation of two ingredients - the heating rate and the radiative transfer. The latter is highly uncertain since it is based on the poorly constrained frequency dependent opacity of r-process elements \citep{Kasen2013,Tanaka2017}. The heating rate, however, can be calculated much more accurately, since it depends mostly on physical properties of r-process elements that are measured by experiments. Here we revisit the thermalization processes with the goal of improving the accuracy of estimates of the heating rate for a given composition and outflow structure. We improve the estimates of the thermalization of electrons by taking into account their experimentally measured initial  energy distribution, which makes a significant difference at late times. The $\gamma$-ray thermalization calculations are also improved by taking into account the $\gamma$-ray spectrum of each decay. Moreover, previous estimates of the opacity for $\gamma$-rays, were based on results obtained for type Ia supernovae (SNe), where the ejecta is dominated by iron-peak elements. Here we take into account the opacity of r-process elements for the $\gamma$-rays that are produced during their decay. We show that this opacity can be  significantly higher than that of iron-peak elements. We develop a publicly available numerical code\footnote{\url{https://github.com/hotokezaka/HeatingRate}} that calculates the heating rate for ejecta with a given composition, mass and velocity profile. Based on this model we also provide an analytic approximated model.

Analytic macronova models are often used  to estimate the ejecta mass and  composition from  observed light curves \citep{li1998ApJ,Grossman2014,Perego2017,Metzger2017LRR,Villar2017,waxman2018}. However, the analytic modelings in the literature oversimplify the photon diffusion process in a homologously expanding ejecta. Either the time delay between the photon production and photon emergence or the effect of the velocity gradient are not taken into account. We present here an analytic modeling which is capable to account for the both effects. 
Comparing the observations and modelings of the macronova light curves, the total mass of r-process elements produced in GW170817  is estimated as $\sim 0.05M_{\odot}$ \citep{Cowperthwaite2017,Drout2017,Kasliwal2017,Perego2017, kasliwal2019,Tanaka2017,Rosswog2018,waxman2018}.
However, this mass estimate involves many systematic uncertainties, mostly due to the uncertain opacity and to lesser extent due to the unknown heating rate and ejecta geometry (see, e.g., \citealt{kasliwal2019,kawaguchi2018}). 
\cite{katz2013} show that a time-weighted integral of the bolometric light curve provides an estimate of the $^{56}$Ni mass produced in supernovae, which depends only on the heating rate and is completely independent of the uncertain radiative transfer. The same method can be used to estimate the total ejecta mass of r-process elements from macronova light curves.
We use our model of the heating rate and apply this method to the GW170817 macronova to obtain robust limits on the total mass of r-process elements produced. 

The structure of the paper is as follows. In \S 2, we introduce decay modes relevant to the macronova heating rate.  We describe the thermalization  processes of decay products in \S 3. We develop an analytic model to calculate the macronova light curve and then compare light curve models with the observed data of the macronova associated with GW170817 in \S 4. We use Katz integral to estimate the ejecta mass in \S 5. We discuss the elemental abundance pattern of merger ejecta in \S 6 and summarize our results in \S 7.

\section{Radioactive power}
R-process nuclei freshly synthesized in neutron star merger ejecta are initially unstable and are disintegrated through $\beta$-decay, $\alpha$-decay, and fission.
In this work, we distinguish the radioactive power from the heating rate. The former describes the radioactive energy generation rate in electrons, $\gamma$-rays, $\alpha$-particles, and fission fragments. The latter means the energy deposition rate of the kinetic energy of decay products to the thermal energy of the ejecta. 
In this section, we describe the radioactive power of $\beta$-decay, $\alpha$-decay, and spontaneous fission. 
\subsection{Beta-decay}
\begin{figure}
\includegraphics[scale=0.55]{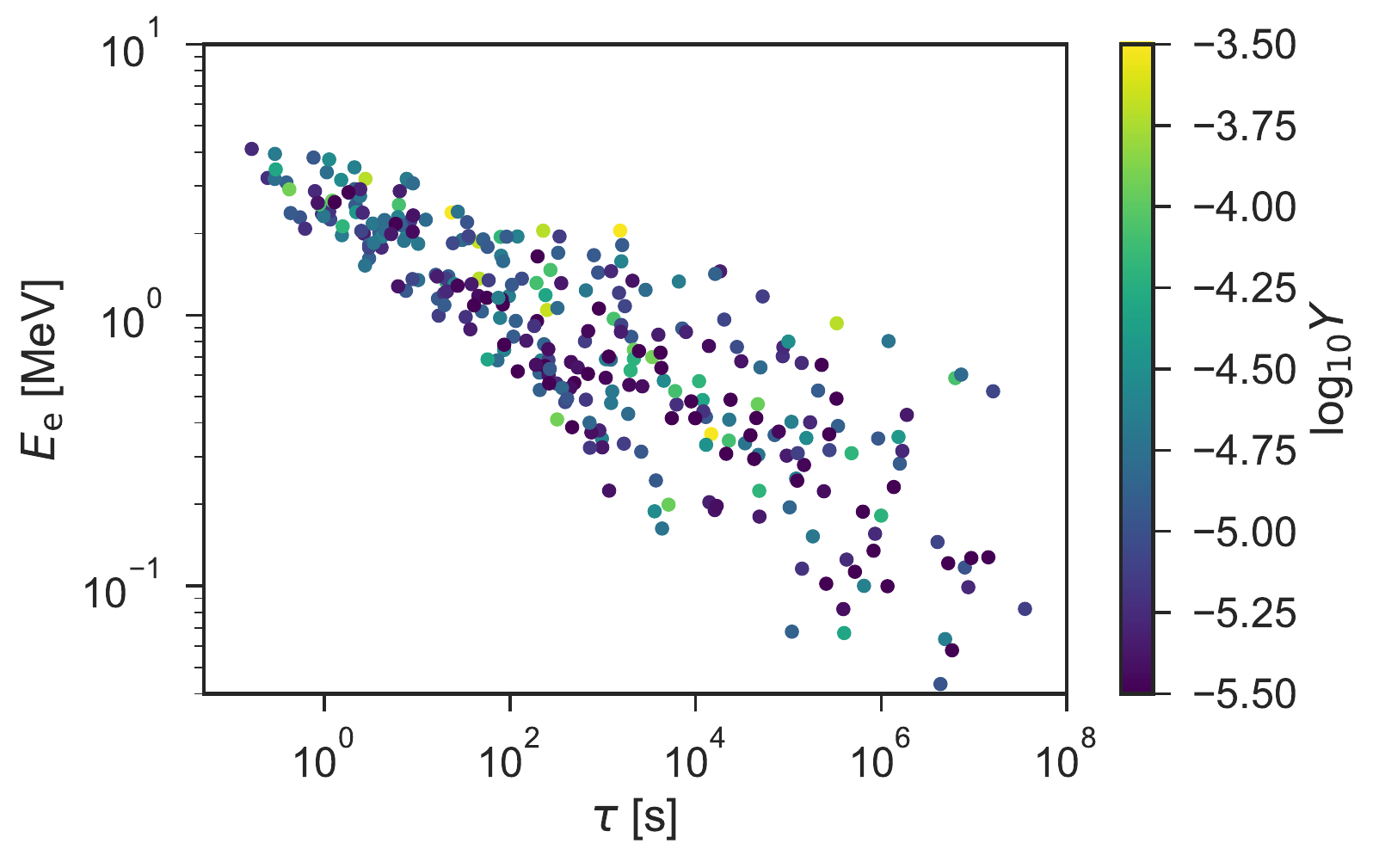}
\includegraphics[scale=0.55]{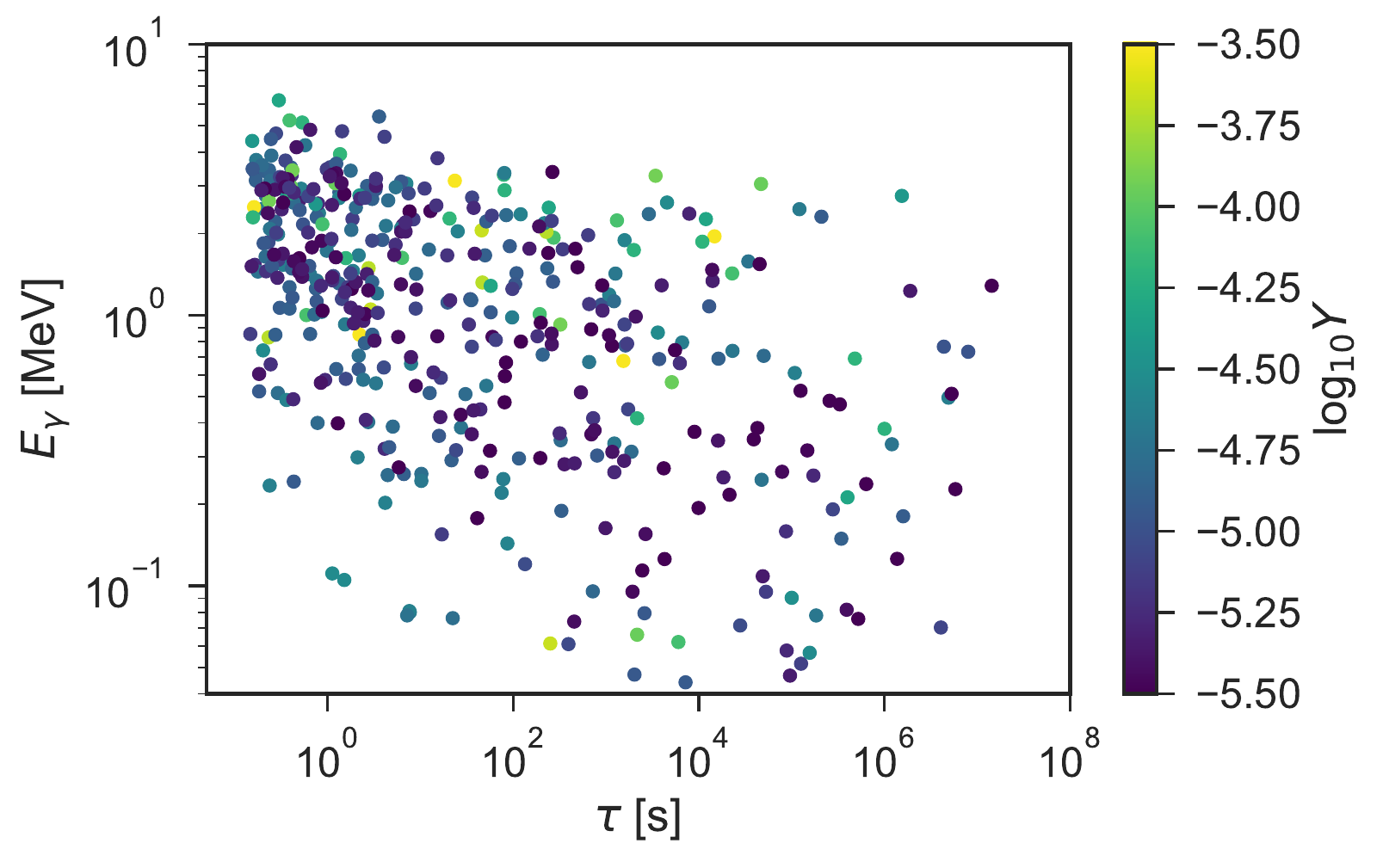}
\caption{
Mean electron energy ({\it left}) and mean $\gamma$-ray energy ({\it right}) released in each $\beta$-decay as a function of mean life-time.  Here  elements with an 
atomic mass number $A\geq 85$ are included. The color of each point shows the solar abundance of r-process elements.
Data are taken from 
Evaluated Nuclear Data File library (ENDF/B-VII.1, \citealt{ENDF}). \vspace{0.5cm}}
\label{fig:beta}
\end{figure}

Majority of r-process nuclei are initially in the $\beta$-unstable region. In particular, those with atomic mass numbers  of $A \leq 209$ undergo $\beta$-decay and approach the stability valley.
The lifetimes of beta-unstable nuclides generally increase as they approach the stability valley.
Each $\beta$-decay releases energy of $\sim 0.1$--$10$ MeV in a neutrino, an electron, and often $\gamma$-rays, where only the electrons and $\gamma$-rays are relevant to the macronova heating rate. 
The total $\beta$-decay energy release in electrons and $\gamma$-rays  are approximated by  $\dot{Q}_{\beta} \approx \dot{Q}_0 t_{\rm day}^{-\delta}$ \citep{Metzger2010,korobkin2012MNRAS,hotokezaka2017MNRAS}. This power-law behavior  is a consequence of the fact that  radioactive decay of many $\beta$-unstable nuclides with different lifetimes contributes to the heat source at different times. Therefore the dependence on the exact composition is not very strong. Nevertheless, the value of $\dot{Q}_0$ can vary with composition by about an order of magnitude were $\dot{Q}_0 \sim 10^{10} {\rm erg/s/g}$  is a typical value, and $\delta$ is typically in the range of $-1.1$ to $-1.4$. For solar abundance pattern with $A\geq 85$ a good analytic approximation of the radioactive power per unit of mass by electrons and $\gamma$-rays is
\begin{eqnarray}
\dot{Q}_{\beta,e}(t) & \approx & 4\times 10^{9}t_{\rm day}^{-1.3} \,{\rm erg/s/g},\\
\dot{Q}_{\beta,\gamma}(t) & \approx & 8\times 10^{9}t_{\rm day}^{-1.4} \,{\rm erg/s/g}.
\end{eqnarray}
Since the radioactive power of the actual elemental abundances of merger ejecta may deviate from the above approximation, in our numerical code  we solve the time evolution of $\beta$-decay chains to get the radioactive power of each decay chain.
We use Evaluated Nuclear Data File library (ENDF/B-VII.1, \citealt{ENDF}) for the injection energies and lifetimes of $\beta$-decays relevant to macronva heating rate (half lives longer than $0.1$\,s).  Figure \ref{fig:beta} shows the half-life and mean energy of electrons and $\gamma$-rays for $\beta$-decay. One can clearly see that the mean electron energy of $\beta$-decay decreases with mean life.

\subsection{Alpha-decay}
R-process nuclei with $209<A\lesssim 250$ may be disintegrated via $\alpha$-decay and end up as   $^{206}$Pb, $^{207}$Pb,  $^{208}$Pb or $^{209}$Bi.
Each $\alpha$-decay releases energy of $\sim 5$--$10$ MeV in the kinetic energy of an $\alpha$-particle. Since  the lifetimes of $\alpha$-unstable nuclides with a larger atomic mass number are typically longer, the first $\alpha$-decay of decay chains may act as a bottleneck. Thus, the first $\alpha$-decay is often followed immediately by several $\alpha$ and $\beta$-decays. \cite{wu2019} show that the $\alpha$-decay chains of $^{222}$Rn ($3.8\,{\rm day},\,23.8\,{\rm MeV}$), $^{223}$Ra ($11.4\,{\rm day},\,30.0\,{\rm MeV}$), $^{224}$Ra ($3.6\,{\rm day},\,30.9\,{\rm MeV}$), and $^{225}$Ra ($14.9\,{\rm day},\,0.4\,{\rm MeV}$) $\rightarrow$ $^{225}$Ac ($10.0\,{\rm day},\,30.2\,{\rm MeV}$), where the half-life and the total energy release per decay chain are shown in the parentheses, are particularly important for the macronova heating rate.  The radioactive power of each decay chain can be approximately estimated as
\begin{eqnarray}
\dot{Q}_{\alpha}(t) \approx 4\cdot 10^{8}e^{-t/\tau} 
\left( \frac{Y_{\alpha}}{10^{-5}}\right)
\left(\frac{\tau}{10\,{\rm day}}\right)^{-1}
\left(\frac{E_{\rm \alpha,tot}}{30\,{\rm MeV}}\right)
\,{\rm erg/s/g},
\end{eqnarray}
where $\tau$ is the mean life, $E_{\rm \alpha,tot}$ is the total energy release per decay chain, $Y_{\alpha}$ is the initial number fraction of a parent nuclide per mole. We take the mean-lives, Q-values, and branching ratios from  ENDF/B-VII.1 to calculate the radioactive power of decay chains.


\subsection{Spontaneous fission}
Finally, transuranium nuclei may be disintegrated via spontaneous fission, which releases large amounts of energy, $\sim 200$ MeV, per decay. Although the abundance of such elements produced in merger ejecta is highly uncertain due to the lack of experimental data, spontaneous fission potentially contributes to the  heating rate \citep{wanajo2014ApJ,Hotokezaka2016MNRAS,barnes2016}.  The radioactive power of spontaneous fission is roughly estimated as
\begin{eqnarray}
\dot{Q}_{\rm sf}(t) \approx 3\cdot 10^{8}e^{-t/\tau} 
\left( \frac{Y_{\rm sf}}{10^{-6}}\right)
\left(\frac{\tau}{10\,{\rm day}}\right)^{-1}
\left(\frac{E_{\rm \rm sf}}{200\,{\rm MeV}}\right)
\,{\rm erg/s/g},
\end{eqnarray}
where $E_{\rm sf}$ is the energy release per spontaneous fission, $Y_{\rm sf}$ is the initial number fraction of a parent nuclide per mole. For instance, the spontaneous fission of $^{254}$Cf is suggested as a possible energy source of macronovae at later times \citep{wanajo2014ApJ,zhu2018,wu2019}. In addition, \cite{wanajo2014ApJ} suggest that $^{259}$Fm and $^{262}$Fm significantly contribute to the heating rate. In later sections, we consider spontaneous fission of $^{254}$Cf, neglecting the minor contribution of $\beta$-decays of the daughter nuclei following fission.

\section{Thermalization}
\subsection{Charged Particles}

Fast-moving charged particles produced by radioactive decay deposit their kinetic energy to the ejecta thermal energy through collisional ionization and excitation, and Coulomb collision with thermal electrons.  
At early times the density of the expanding ejecta is high enough that the collisional energy loss occurs on time scales shorter than one dynamical time. 
In this regime, the  heating rate is practically the same to the radioactive power at any given time. At later times, however, collisional thermalization takes longer time than one dynamical time.   As a result, the heating rate deviates from the  radioactive power at a given time \citep{barnes2016,kasen2018,waxman2019}.  Our calculation is similar to the analytic methods presented by \cite{kasen2018,waxman2019}, but we specify the injection energies  of decay products for each decay chain, which as we show below can have a significant effect on the thermalization efficiency at late times. Note, that these energies are known rather well since they at the relevant times ($t \gtrsim 10^3$ s), all unstable nuclides are very close to the valley of stability where there is direct experimental data.

The collisional energy loss of decay products per unit time is  described by  $K_{\rm st}v\rho$, where $K_{\rm st}$ is the stopping cross section per unit mass in units of ${\rm MeV\,cm^2/g}$, $v$ is the velocity of a fast particle, and $\rho$ is the density of the stopping medium (see Appendix A). 
Figure \ref{fig:stop_ele} shows $\beta K_{\rm st}$ for electrons and $\alpha$-particles, where $\beta$ is the velocity normalized by the speed of light. This quantity is a proxy of the energy loss rate of fast particles. The stopping power due to ionization and excitation 
 peaks around the energy at which the velocity of a
fast particle is approximately the orbital velocity of an atomic electron with the mean binding energy $\langle I \rangle\approx 500Z_{50}$ eV corresponding to $0.05cZ_{50}^{1/2}$, where $Z_{50}$ is the atomic number of the stopping medium normalized by $50$.
For $\beta$-decay electrons, their initial velocities are always much faster than this velocity so that $\beta K_{\rm st}$ increases as they lose energy. 
As one can see in figure \ref{fig:stop_ele},  $\beta K_{\rm st}$ for electrons  increases very slowly with decreasing  energy. For $\alpha$-particles, they are injected around the peak of the ionization stopping power  and the stopping power of thermal electrons starts to dominate below $0.1$--$1$ MeV. Consequently, $\beta K_{\rm st}$ for $\alpha$-particles also slowly increases in the energy loss process.  The roughly flat spectrum of $\beta K_{\rm st}$ of electrons and $\alpha$-particles means that  the fractional energy loss due to collision occurs faster for particles with lower energy. On the contrary,  the initial velocities of fission fragments are typically slower than the orbital velocity of atomic electrons. The ionization stopping power behaves as $\beta K_{\rm st}\propto E$ for  energies down to $\sim 10$ MeV, where the stopping power of thermal electrons becomes more important (see  
figure \ref{fig:stop_sf}), implying that the fractional energy loss of fission fragments with different energies occurs with roughly the same rate.
Note that the dependence of $\beta K_{\rm st}$ on the atomic number of the stopping medium is rather weak. For instance, the difference in $\beta K_{\rm st}$ between Xe and U is  $\sim 20\%$ for electrons and $\sim 50\%$ for 
$\alpha$-particles. Thus, we use $\beta K_{\rm st}$  of Xe in the following calculations.

For comparison of  electron thermalization to that of $\alpha$-particles and fission fragments, it is useful to define an effective opacity as $\kappa_{\rm eff}=\beta K_{\rm st}/E$. With this definition a charged particle deposits a significant fraction of its energy after spending one dynamical time $t$ in ejecta when the effective optical depth is $\tau_{\rm eff}= \kappa_{{\rm eff}}\rho ct\gtrsim 1$. The effective opacity of $\beta$-electrons, $\alpha$-particles, and fission fragments is
\begin{eqnarray}
\kappa_{\beta,{\rm eff}} & \approx & 4.5\,{\rm cm^2/g}\left(\frac{E}{0.25\,{\rm MeV}}\right)^{-1},\\
\kappa_{\alpha,{\rm eff}} & \approx & 3\,{\rm cm^2/g}\left(\frac{E}{7\,{\rm MeV}}\right)^{-1},\\
\kappa_{\rm sf,eff} & \approx & 10\,{\rm cm^2/g},
\end{eqnarray}
respectively. 

The thermalization time is defined by the time where the effective optical depth is unity. For electrons it is
\begin{eqnarray}
t_{{\rm th},\beta} &\approx & \left(\frac{C_{\rho}c  \kappa_{\beta,{\rm eff}} M_{\rm ej}}{ v_0^3} \right)^{1/2}\\
& \approx & 55\,{\rm day}\,\left(\frac{C_{\rho}}{0.05}\right)^{1/2}
\left(\frac{M_{\rm ej}}{0.05M_{\odot}}\right)^{1/2}
\left(\frac{\kappa_{\beta,{\rm eff}}}{4.5\,{\rm cm^2/g}}\right)^{1/2}
\left(\frac{v_{0}}{0.1c}\right)^{-3/2}.
\label{eq:t_th_beta}
\end{eqnarray}
For $\alpha$-particles
\begin{eqnarray}
t_{\rm th,\alpha} \approx 45\,{\rm day}\,
\left(\frac{C_{\rho}}{0.05}\right)^{1/2}
\left(\frac{M_{\rm ej}}{0.05M_{\odot}}\right)^{1/2}
\left(\frac{\kappa_{\alpha,{\rm eff}}}{3\,{\rm cm^2/g}}\right)^{1/2}
\left(\frac{v_{0}}{0.1c}\right)^{-3/2},
\end{eqnarray}
and for fission fragments,
\begin{eqnarray}
t_{\rm th,sf} \approx 85\,{\rm day}\,\left(\frac{C_{\rho}}{0.05}\right)^{1/2}
\left(\frac{M_{\rm ej}}{0.05M_{\odot}}\right)^{1/2}
\left(\frac{\kappa_{{\rm sf,eff}}}{10\,{\rm cm^2/g}}\right)^{1/2}
\left(\frac{v_{0}}{0.1c}\right)^{-3/2},
\end{eqnarray}
where $M_{\rm ej}$ is the ejecta mass, $v_0$ is the minimum ejecta velocity, and $C_{\rho}$ is a coefficient that depends on the ejecta velocity profile. Note that for electrons and $\alpha$-particles there is an implicit dependence on the particle injection energy via $\kappa_{\rm eff}$ such that $t_{\rm th} \approxprop E^{-1/2}$.

In this work,  we use a radial density profile of merger ejecta:
\begin{eqnarray}
  \rho(t,v) = 
        \rho_0(t)\left(\frac{v_{\rm ej}}{v_0}\right)^{-n} ~~~~(v_0\leq v_{\rm ej}\leq v_{\rm max})\label{eq:density}
 \end{eqnarray}
 where $\rho_0(t)$ is defined such that 
 \begin{eqnarray}
 M_{\rm ej} = 4\pi \int_{v_0}^{v_{\rm max}}dvv^2 \rho(t,v).
 \end{eqnarray}
The density profile corresponds to the mass profile $dm/d\ln v\propto v^{-k}$, where $k=n-3$. Under the most reasonable assumption that the fast particles are trapped within the ejecta by random magnetic fields, $C_\rho$ can be approximated using the mass weighted density:
\begin{eqnarray}
 \rho_m(t) = \frac{\int dm \rho}{M_{\rm ej}} = C_{\rho} M_{\rm ej}v_0^{-3}t^{-3},
\end{eqnarray}
which for the power-law profile we consider gives\footnote{The thermalization time of electrons obtained by plugging equation (\ref{eq:Crho}) into equation (\ref{eq:t_th_beta}) is similar to the one obtained by \cite{waxman2018}, with one difference. \cite{waxman2018} does not consider the maximal velocity of the ejecta and therefore they are missing the term $(1-w^k)^2$ in the denominator of equation (\ref{eq:Crho}). This term may be important when the velocity distribution is flat, i.e., $k < 1$. }
\begin{eqnarray}
C_{\rho} \approx \frac{k}{4\pi(2+3/k)(1-w^k)^2},
\label{eq:Crho}
\end{eqnarray}
where $w=v_{0}/v_{\rm max}$, e.g., $C_{\rho}\approx 0.03$  for $n=4$ ($k=1$)  and $w= 0.25$.

The thermalization time, $t_{\rm th}$, is the characteristic time at which thermalization becomes inefficient. An accurate calculation of the thermalization at $t \gtrsim t_{\rm th}$  requires following the time evolution of the kinetic energy of monoenergetic charged particles, which is solved by
\begin{eqnarray}
\frac{dE}{dt} = -K_{\rm st}\rho_m v_c -3(\gamma_{\rm ad}-1)\frac{E}{t},\label{eq:cool}
\end{eqnarray}
where   $\gamma_{\rm ad}$ is the adiabatic index of charged particles.  
The value of the adiabatic index, $\gamma_{\rm ad}$, depends on the type of decay products as well as energy.  $\alpha$-particles and fission fragments are always non-relativistic; therefore,   $\gamma_{\rm ad}$  is $5/3$. For $\beta$-decay electrons, $\gamma_{\rm ad}$ varies in between $5/3$ and $4/3$
because the initial kinetic energy ranges from $\approx 0.1$ to a few MeV. 
The adiabatic index  of  monoenergetic electrons  is given by (e.g., \citealt{nakar2008})
\begin{eqnarray}
\gamma_{\rm ad}(p) = 1+\frac{p^2}{3\sqrt{p^2+1}(\sqrt{p^2+1}-1)},
\end{eqnarray}
where $p$ is the electron's momentum in units of $m_e$ and $c$. In the calculation presented later, we use a constant value of $\gamma_{\rm ad}$ at an initial momentum of  $\beta$-electrons  (see Appendix B for the approximated solution of equation \ref{eq:cool}).


\begin{figure}
\begin{center}
\includegraphics[scale=0.35]{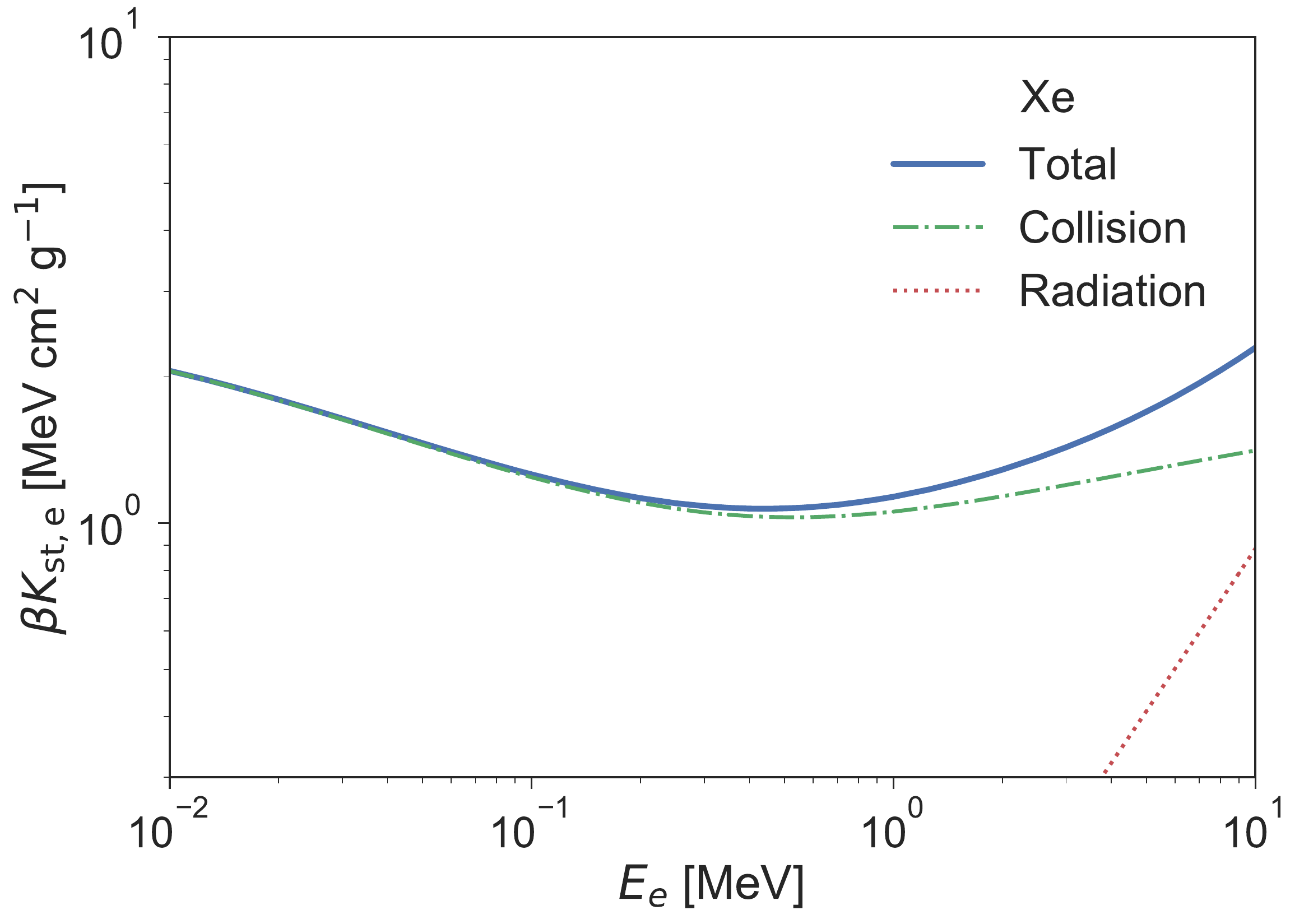}
\includegraphics[scale=0.35]{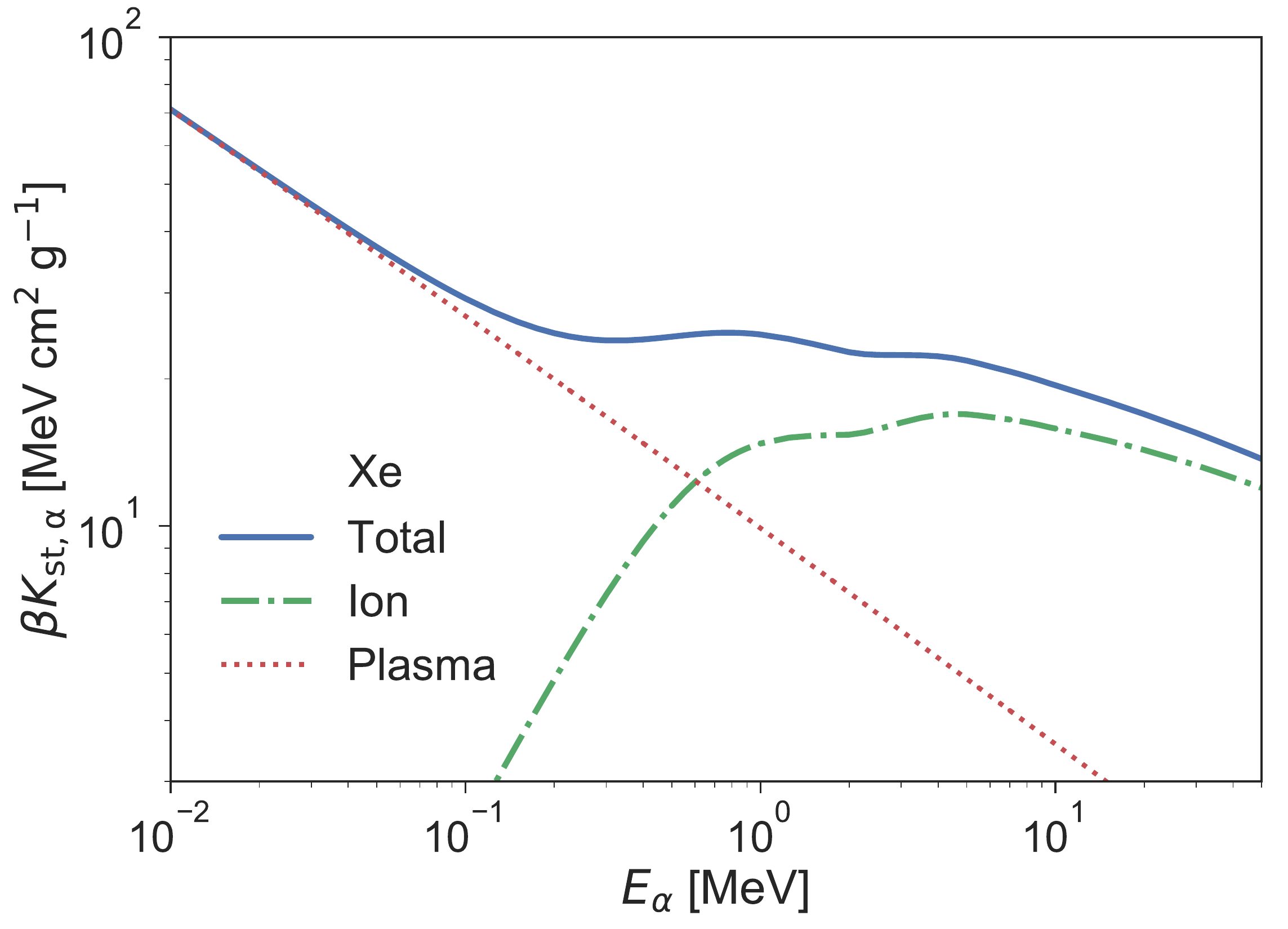}
\end{center}
\caption{
Stopping power for electrons and $\alpha$-particles. Here the stopping medium is chosen to be xenon, which is in the second r-process peak, as an example.
The stopping power due to ionization and excitation is taken from ESTAR and ASTAR of NIST database (\url{http://physics.nist.gov/Star}) for electrons and $\alpha$-particles, respectively. The stopping power due to Bremsstrahlung is also shown for electrons. For $\alpha$-particles, the contribution of the Coulomb collision with thermal electrons is calculated by using the Bohr's formula \citep{bohr}. Here we assume xenon is singly ionized. \vspace{0.5cm}
}
\label{fig:stop_ele}
\end{figure}

\begin{figure*}
\begin{center}
\includegraphics[scale=0.35]{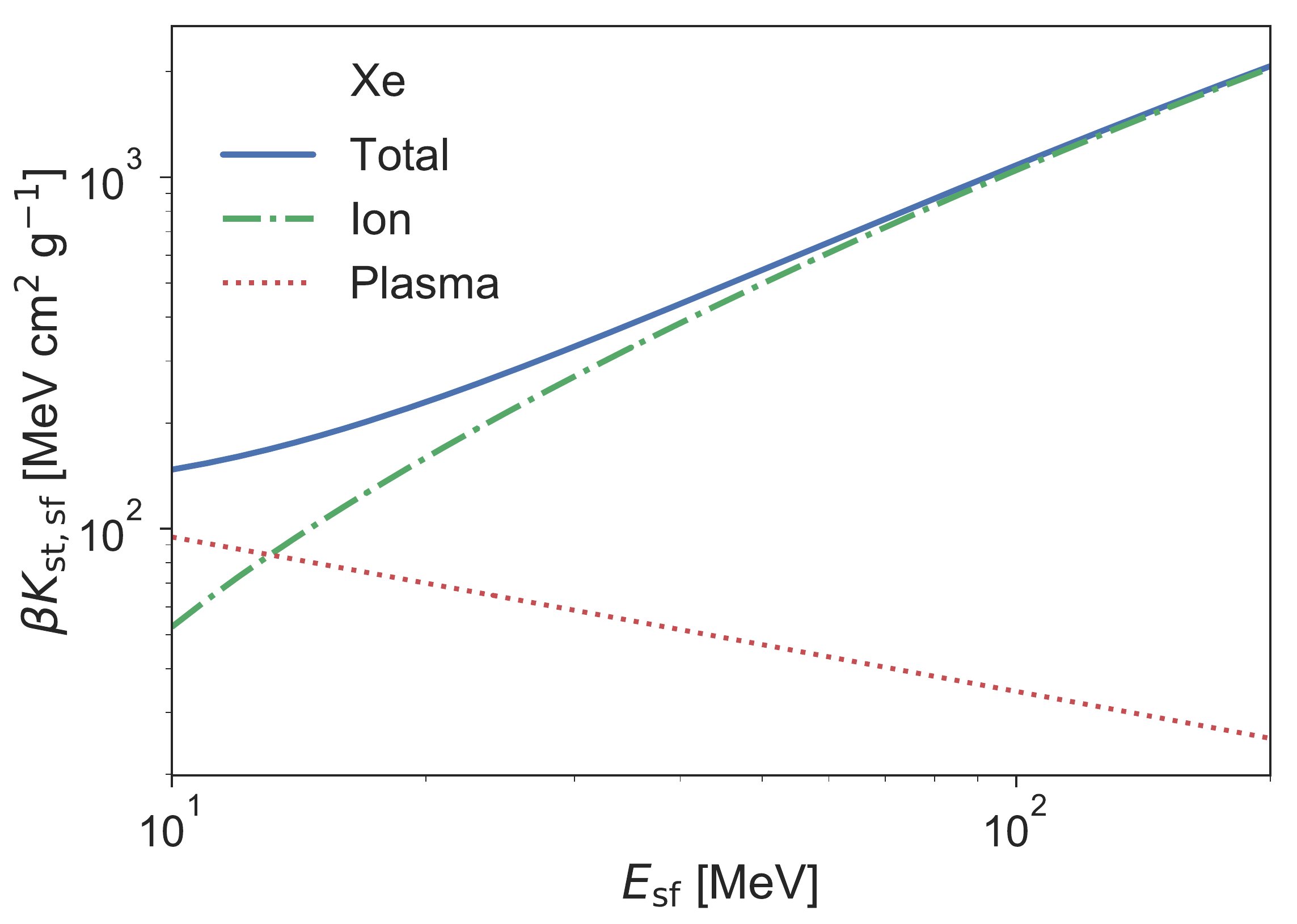}
\end{center}
\caption{
Same as figure \ref{fig:stop_ele} but for fission fragments \citep{Mukherji1974}.
}
\label{fig:stop_sf}
\end{figure*}

For $t\gtrsim t_{\rm th}$,  charged particles do not lose their kinetic energy within one dynamical time.
In this phase, one needs to take into account not only particles injected at $t$ but also those injected at earlier times, which may  contribute to or even dominate the heating rate.  The  heating rate at a given time $t$ is obtained by integrating all the contribution of non-thermal particles:
\begin{eqnarray}
\dot{Q}_{\rm th}(t) & = & \sum_i\int^t_{t_{0,i}}dt' \beta K_{\rm st}(E_{i,0};t',t)\rho(t)\frac{N_i(t')}{\tau_i},
\end{eqnarray}
where  $\beta K_{\rm st}(E_{i,0};t',t)$ is obtained by 
solving equation (\ref{eq:cool}) for a given initial energy, $E_{i,0}$, and injected time $t'$, $N(t)$ is the number density of a radioactive element $i$. Here, the lower limit of the integral $t_{0,i}$ corresponds to the time when the oldest non-thermal charged particles surviving at $t$ are produced.

\begin{figure}
\begin{center}
\includegraphics[scale=0.55]{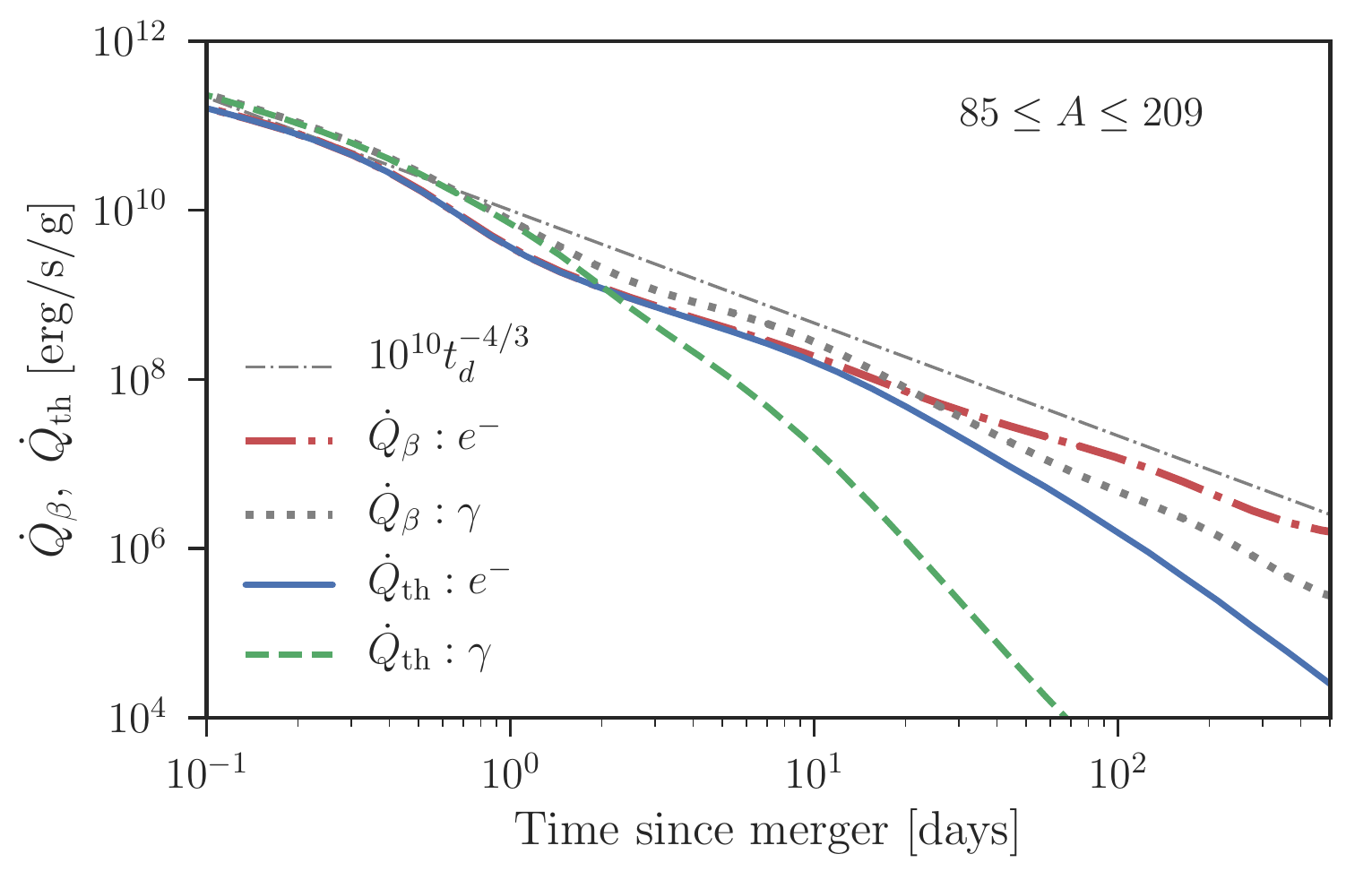}
\includegraphics[scale=0.55]{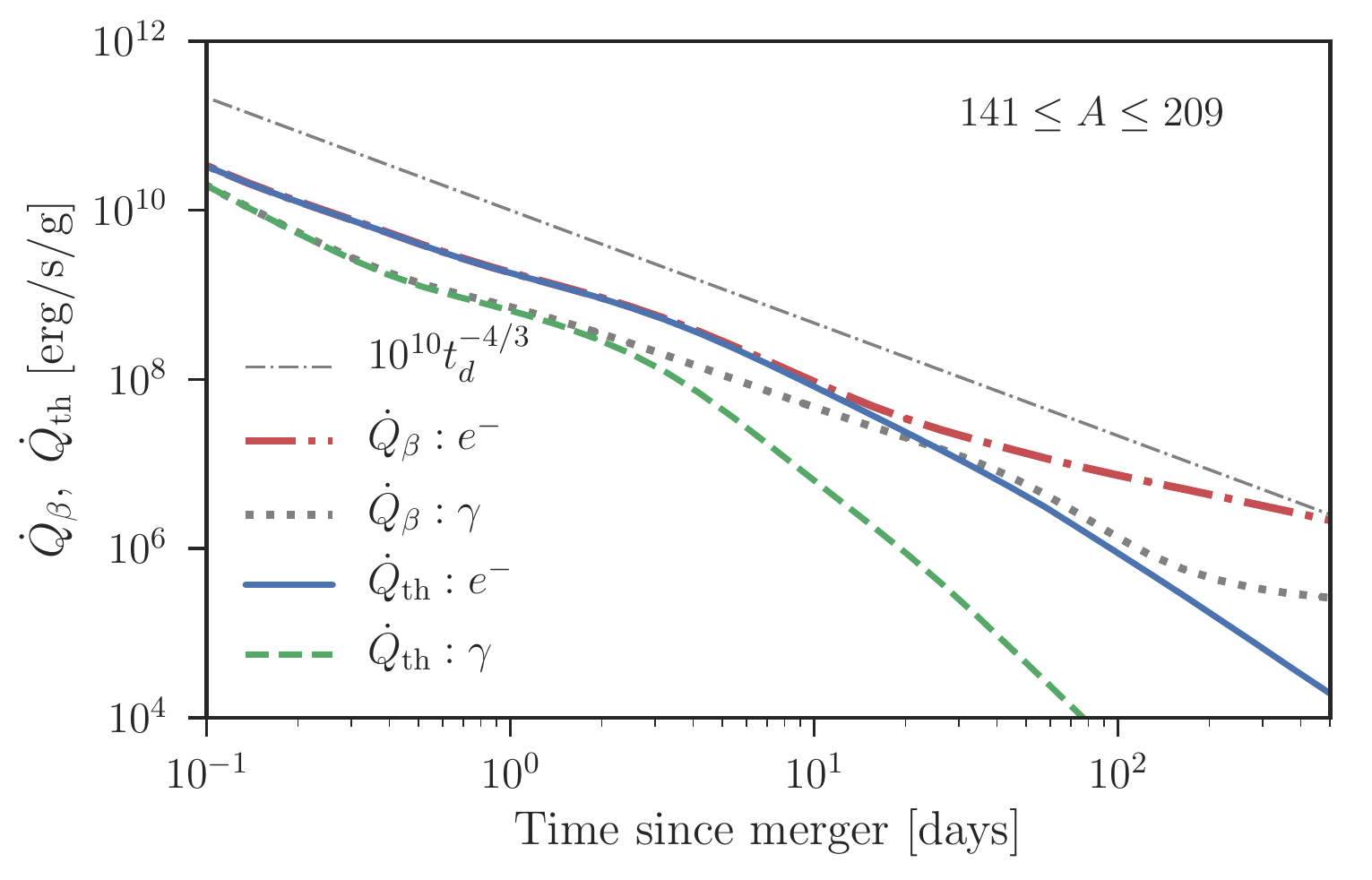}
\end{center}
\caption{
 Radioactive power and heating rate of $\beta$-decay in electrons and $\gamma$-rays. The solar r-process abundance pattern with a minimum atomic mass number of $A_{\rm min}=85$ ({\it left}) and $141$ ({\it right}) is assumed.  Also shown in both panels is an analytic heating rate, $10^{10}(t/{\rm day})^{-4/3}\,{\rm erg/s/g}$.  For the thermalization processes, we assume an ejecta mass of $0.05M_{\odot}$, $v_0=0.1c$, $v_{\rm max}=0.4c$ and $n=4.5$.\vspace{0.5cm}}
\label{fig:h1}
\end{figure}

Figure \ref{fig:h1}  shows the  heating rates of $\beta$-decay. Here, we assume  the solar r-process abundance of $85\leq A\leq 209$ ({\it left}) and $141\leq A\leq 209$ ({\it right}).
The former choice includes the second r-process peak and the latter does not. Including the second r-process peak enhances the heating rate 
and results in more radioactive power in $\gamma$-rays.
The electron heating rates  start to deviate significantly from the radioactive power around 20-30 days after the merger and after 50-80 days they reach an asymptotic decline with a power law $\dot{Q}_{\rm th}\propto t^{-2.8}$. We find that this transition of the heating rate occurs rather slowly compared to the heating rate with the analytic description presented by \cite{barnes2016}.

\begin{figure}
\begin{center}
\includegraphics[scale=0.55]{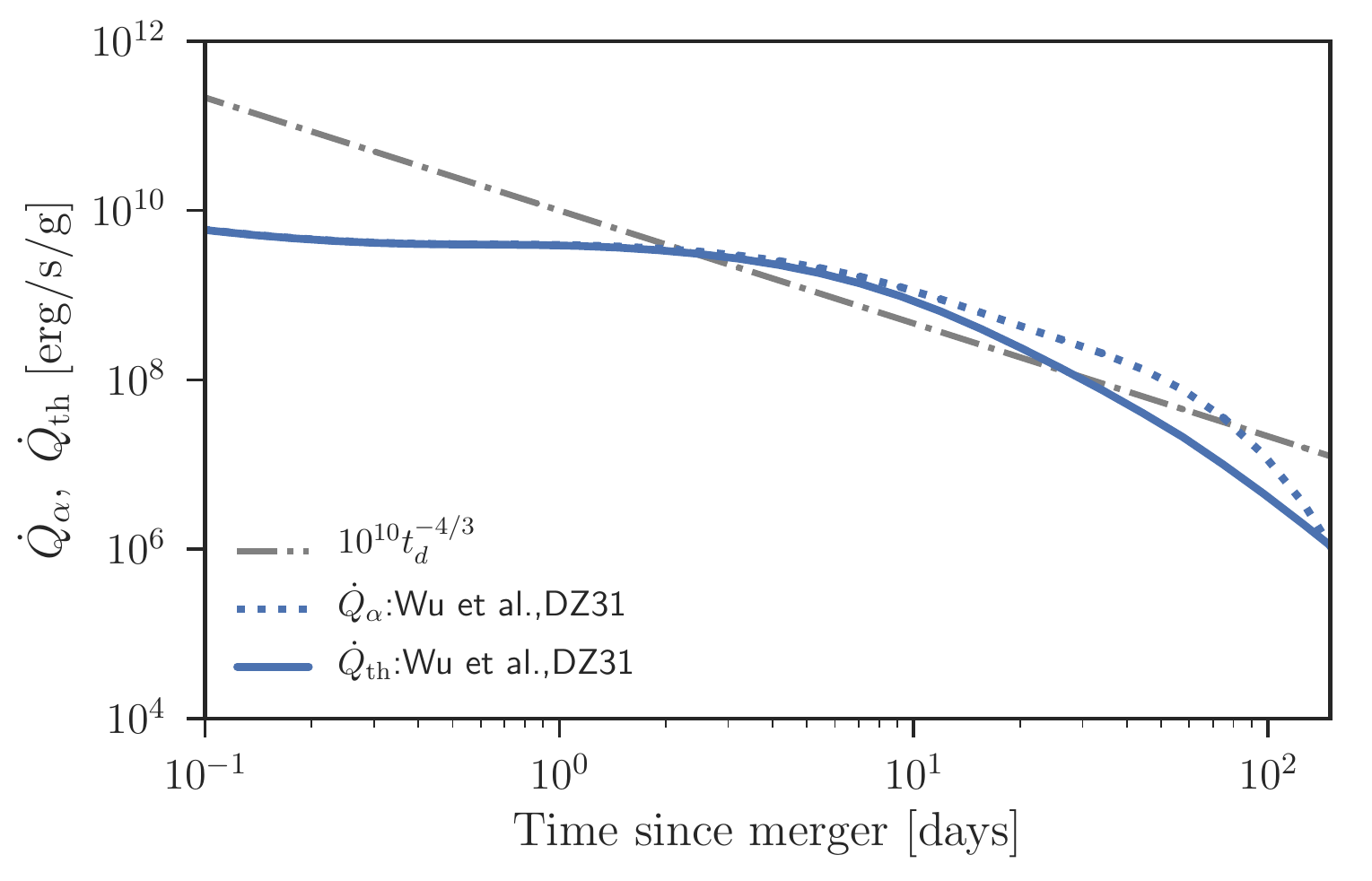}
\includegraphics[scale=0.55]{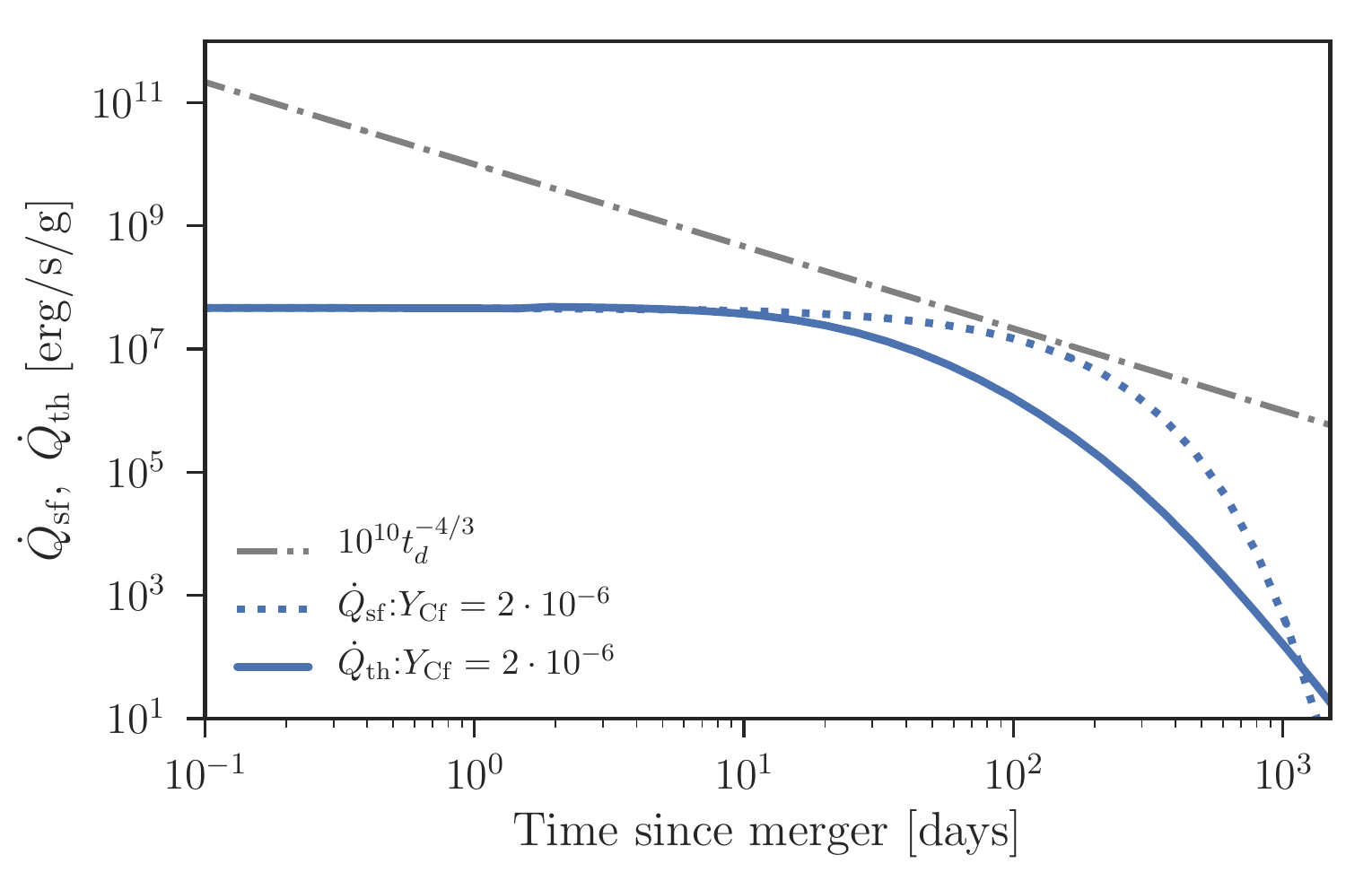}
\end{center}
\caption{
Same as figure \ref{fig:h1} but for $\alpha$-decay ({\it left}) and spontaneous fission ({\it right}). Here we assume  the initial abundance of $\alpha$-decay nuclei of
$(Y(^{222}{\rm Rn}), Y(^{223}{\rm Ra}), Y(^{224}{\rm Ra}), Y(^{225}{\rm Ra}))=(4.0\cdot 10^{-5},2.7\cdot 10^{-5},4.1\cdot 10^{-5},2.7\cdot 10^{-5})$ \citep{wu2019}. $^{254}$Cf with an initial abundance of $2.0\cdot 10^{-6}$  is used. The ejecta profile same to that of figure \ref{fig:h1} is used.\vspace{0.5cm}
}
\label{fig:h2}
\end{figure}

Figure \ref{fig:h2} shows the heating rate and radioactive power of $\alpha$-decay and fission.
For $\alpha$-decay, we use the initial abundances of $\alpha$-decaying nuclei of $222 \leq A\leq 225$ of the DZ31 model \citep{wu2019}. Note that this model predicts the production of particularly large amounts of
these nuclei. The late-time heating rate of $\alpha$-decay approaches $\propto t^{-2.8}$. For spontaneous fission, we consider only $^{254}$Cf and its heating rate declines as $t^{-3}$ around $t\sim \tau$ and $t^{-5}$ for $t\gg \tau$.

This power law behavior at later times $t\gtrsim t_{\rm th}$ is quite general and explained as follows.
The total number of non-thermal charged particles is approximately constant with time for $t > t_{{\rm th}}$, i.e., $\sum_i\int dt' N_i(t')/\tau_i \approx {\rm const}$.
This is true in the case that  many beta-decay chains contribute to the heating rate
as well as in the case that a few decaying species dominate the radioactive power around the thermalization time, i.e, $t\gtrsim \tau \sim t_{\rm th}$.   If one neglects the energy dependence of $\beta K_{\rm st}$, the time dependence of the thermalization rate is simply $\dot{Q}_{\rm th}(t)\propto \rho(t) \propto t^{-3}$. Given  the weak energy dependence of $\beta K_{\rm st}$ for electrons and $\alpha$-particles, the time evolution of the heating rate is approximately described as
\begin{eqnarray}
\dot{Q}_{\rm th}(t) \approxprop t^{-2.8}~~~~~~~(\mbox{$\beta$-decay and $\alpha$-decay}).
\end{eqnarray}
For fission products $\beta K_{\rm st}$ drops as they cool adiabatically and therefore the asymptotic decay of $\dot{Q}_{\rm th}$ is faster. 

Analytic solutions of the $\beta$-decay heating rate of r-process elements are derived by \cite{kasen2018} and \cite{waxman2019}. \cite{kasen2018} find that the heating rate approaches $\approxprop t^{-7/3}$ at late times. In their calculation they neglect the   logarithmic factor of the stopping cross section, i.e., $\beta K_{\rm st,e} \approxprop E^{-0.5}$. \cite{waxman2019} take a proper account of this factor obtaining $\beta K_{\rm st,e} \approxprop E^{-0.15}$. With this energy dependence of the cross-section 
they obtain a steeper decline of the heating rate $\approxprop t^{-2.8}$. We use the complete stopping cross section formula, with the logarithmic factor  (see figure \ref{fig:stop_ele}); therefore, the time dependence of our late-time electron heating rate agrees with that obtained by \cite{waxman2019}.

\cite{waxman2018} interpret a break in the observed bolometric light curve of the GW170817 macronova around $\sim 6$ days as the thermalization break, which is a transition of the electron heating rate from the regime of $\dot{Q}_{{\rm th},e}\propto t^{-1.3}$ to $\propto t^{-2.8}$. This scenario requires that the thermalization break occurs at relatively early time and that it is rather sharp. As one can see in figure \ref{fig:h1}, which is obtained for ejecta that is similar to the one inferred for GW170817, the thermalization break seems to take place on a significantly longer time scale and is too gradual to reproduce the observed break. We discuss  the break in the observed light curve  in \S 4.

\subsection{Gamma-rays}
\begin{figure}
\begin{center}
\includegraphics[scale=0.45]{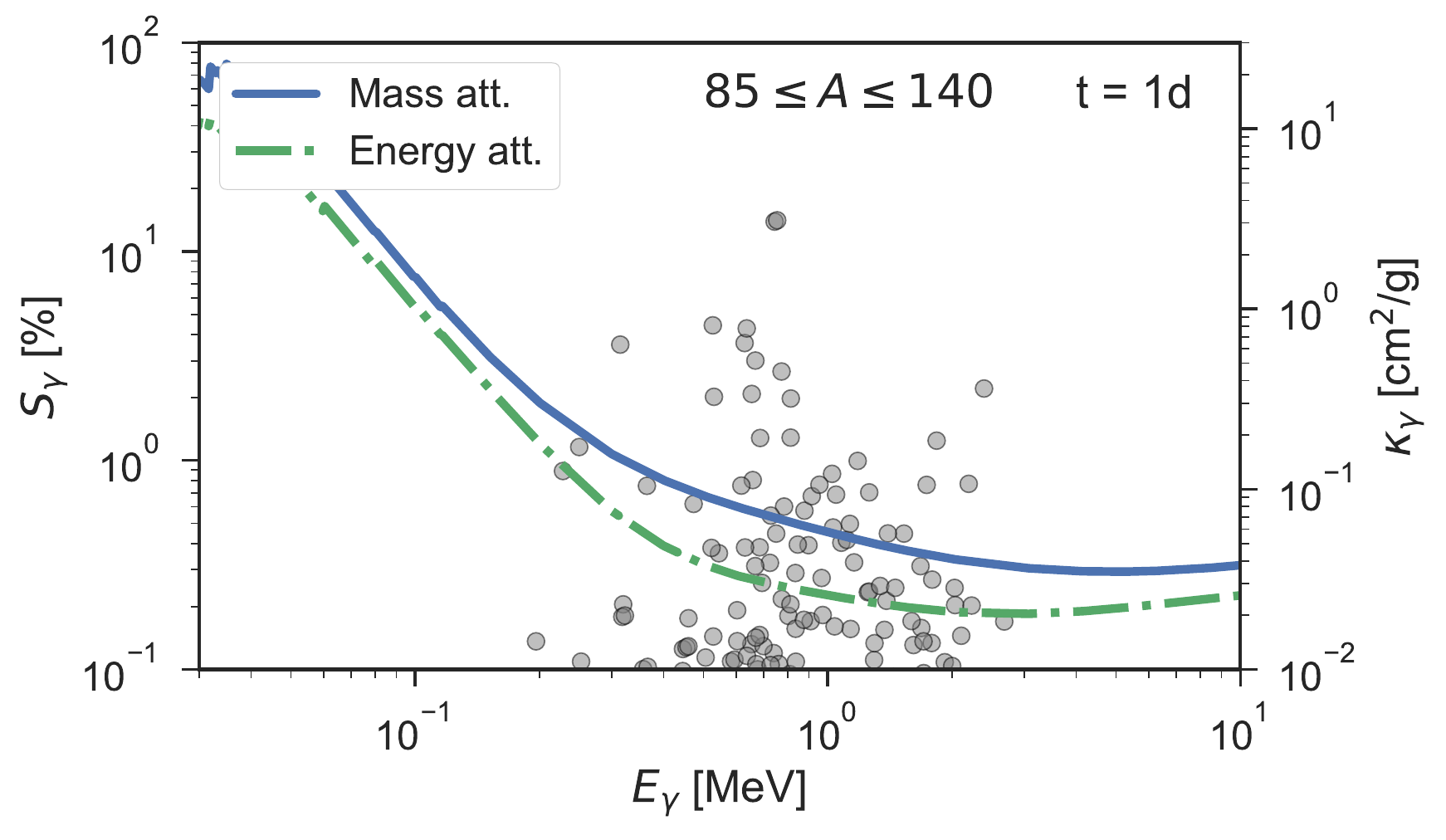}
\includegraphics[scale=0.45]{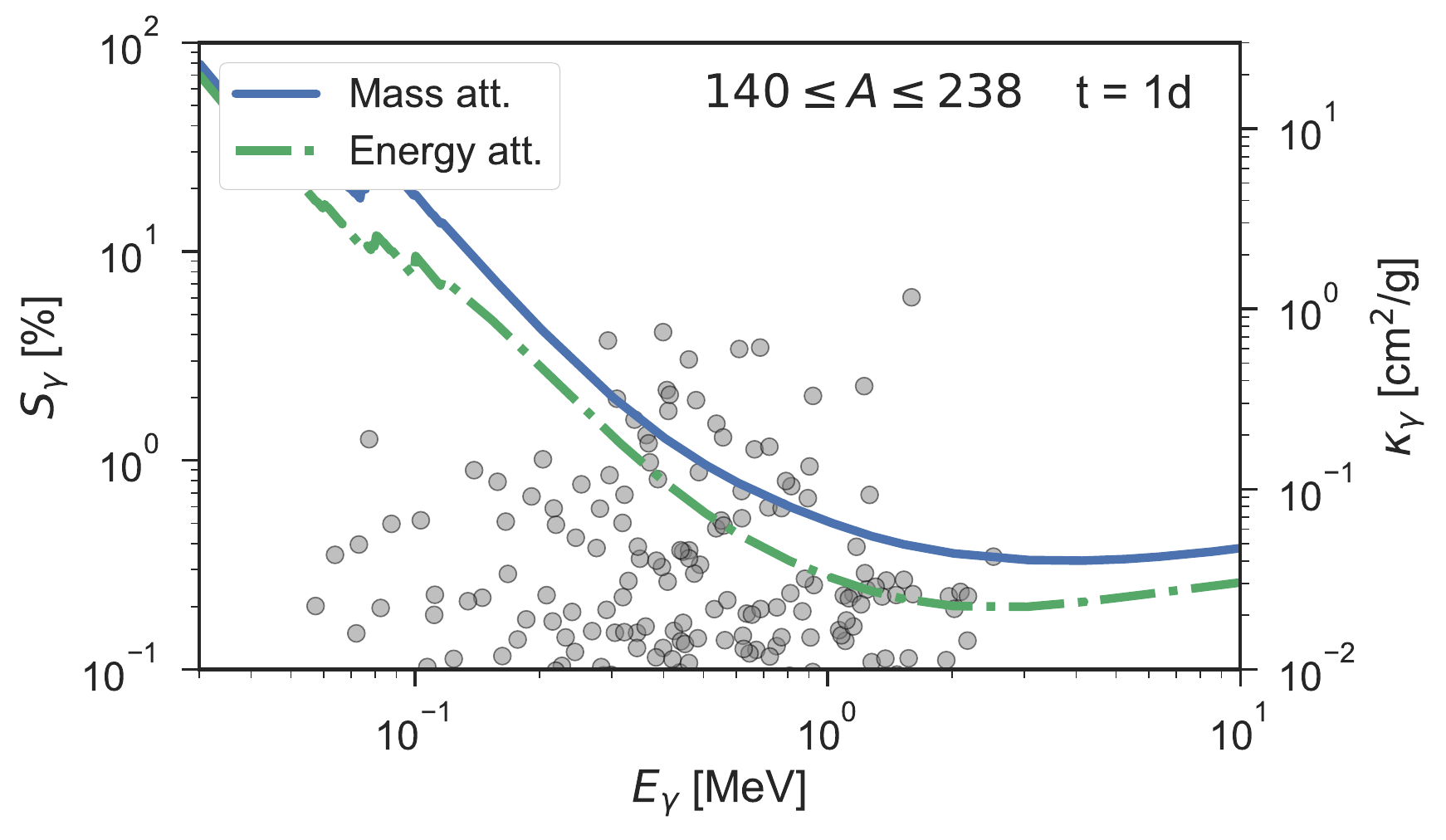}\\
\end{center}
\caption{
$\gamma$-ray spectrum of $\beta$-decay and
mass and energy attenuation coefficients of r-process material 
for $\gamma$-rays. The nuclear abundance pattern is assumed to
be the solar r-process abundance with mass numbers of $85\leq A 140$ ({\it left})
and $140\leq A \leq 238$ ({\it right}). The $\gamma$-ray spectra at 1 day ({\it top})
and 3 day ({\it bottom}) are shown.  \vspace{0.5cm}}
\label{fig:stop_gamma}
\end{figure}

$\gamma$-rays are often produced by radioactive decay and their energy ranges from $\sim 0.1$ to a few MeV.
 These $\gamma$-rays may interact with electrons and deposit their energy to the ejecta's thermal energy through Compton scattering, photoelectric absorption, and pair creation. Figure \ref{fig:stop_gamma} shows the  opacity of r-process elements for $\gamma$-rays. Also shown are the spectral energy distribution of $\gamma$-rays produced by $\beta$-decays. Note that the opacity of heavy material ($140\leq A\leq 238$) is larger by a factor of $\gtrsim 1.5$ than that of lighter elements  at low energies $\lesssim 0.5$ MeV because photoelectric absorption is enhanced for high $Z$ atoms. In addition, $\gamma$-rays are emitted at slightly lower energies for heavier elements. 

Typically $\gamma$-rays first lose their energy through Compton scattering. The down scattered $\gamma$-rays then may be destroyed by photoelectric absorption.  It is not trivial to evaluate the energy deposition fraction of $\gamma$-rays. In the context of SNe Ia, the results of Monte-Carlo simulations of $\gamma$-ray transfer  show that  
the fraction of $\gamma$-rays energy which is deposited to the thermal energy at any given time can be estimated rather accurately by 
finding a time scale $t_0$ \citep{swartz1995,jeffery1999,wygoda2019}. This time scale, $t_0$, is defined by the time
at which the effective optical depth for $\gamma$-rays is unity, $\tau_{\rm \gamma,eff}=1$:
\begin{eqnarray}
\tau_{\rm \gamma,eff} = \kappa_{\rm \gamma,eff}\Sigma_m(t),
\end{eqnarray}
where $\kappa_{\rm \gamma,eff}$ is the purely absorptive effective opacity and
the mass-weighted column density of the ejecta is
\begin{eqnarray}\label{eq:sigma}
\Sigma_m (t) & = & \int d^3x\frac{\rho(t,\vec{x})}{M_{\rm ej}} \int \frac{d\hat{\Omega}}{4\pi} \int_0^{\infty} ds \rho(t,\vec{x}+s\hat{\vec{\Omega}}),\\
& = & C_{\Sigma} M_{\rm ej}v_0^{-2}t^{-2},
\end{eqnarray}
where $\hat{\vec{\Omega}}$ is the unit solid angle vector. $C_{\Sigma}$ is a constant that depend on the structure of the ejecta and can be found by carrying out the integral in the equation \ref{eq:sigma}. For the ejecta power-law profile that we consider (equation \ref{eq:density}) the integration should be carried out numerically. The analytic formula
\begin{eqnarray}
C_{\Sigma} \approx 0.1w+0.003\frac{k}{w}.
\end{eqnarray}
provide a good approximation (up to a factor of order unity) for $0<k<5$ and $0.1<w<0.5$, which is the most relevant range for the merger ejecta.

The effective opacity $\kappa_{\rm \gamma,eff}$ accounts for the fraction of the energy that $\gamma$-rays deposit when they propagate through a unit of column mass density. It is averaged over the $\gamma$-rays produced by each radioactive decay
\begin{eqnarray}\label{eq:kappa_gamma}
\kappa_{\rm \gamma,eff} = \sum_i f_{\gamma,i}\kappa_{\gamma}(E_i),
\end{eqnarray}
where $f_{\gamma,i}$ is the fraction of energy emitted in the $\gamma$-ray line $i$ of energy $E_i$ and the opacity term $\kappa_{\rm \gamma}(E_i)$ is approximated by the geometric mean of the mass and the energy attenuation opacities at $E_i$. 
The values of $\kappa_{\rm \gamma,eff}$ are typically in a range from $\approx 0.02$ to $\approx 1\,{\rm cm^2/g}$ depending on the $\gamma$-ray spectrum.  
As we noted above, $\kappa_{\rm \gamma,eff}$ of heavy elements is larger than lighter elements and therefore an accurate calculation requires to specify $\kappa_{\rm \gamma, eff}$ for each decay for a given ejecta composition.  

\begin{figure}
\begin{center}
\includegraphics[scale=0.5]{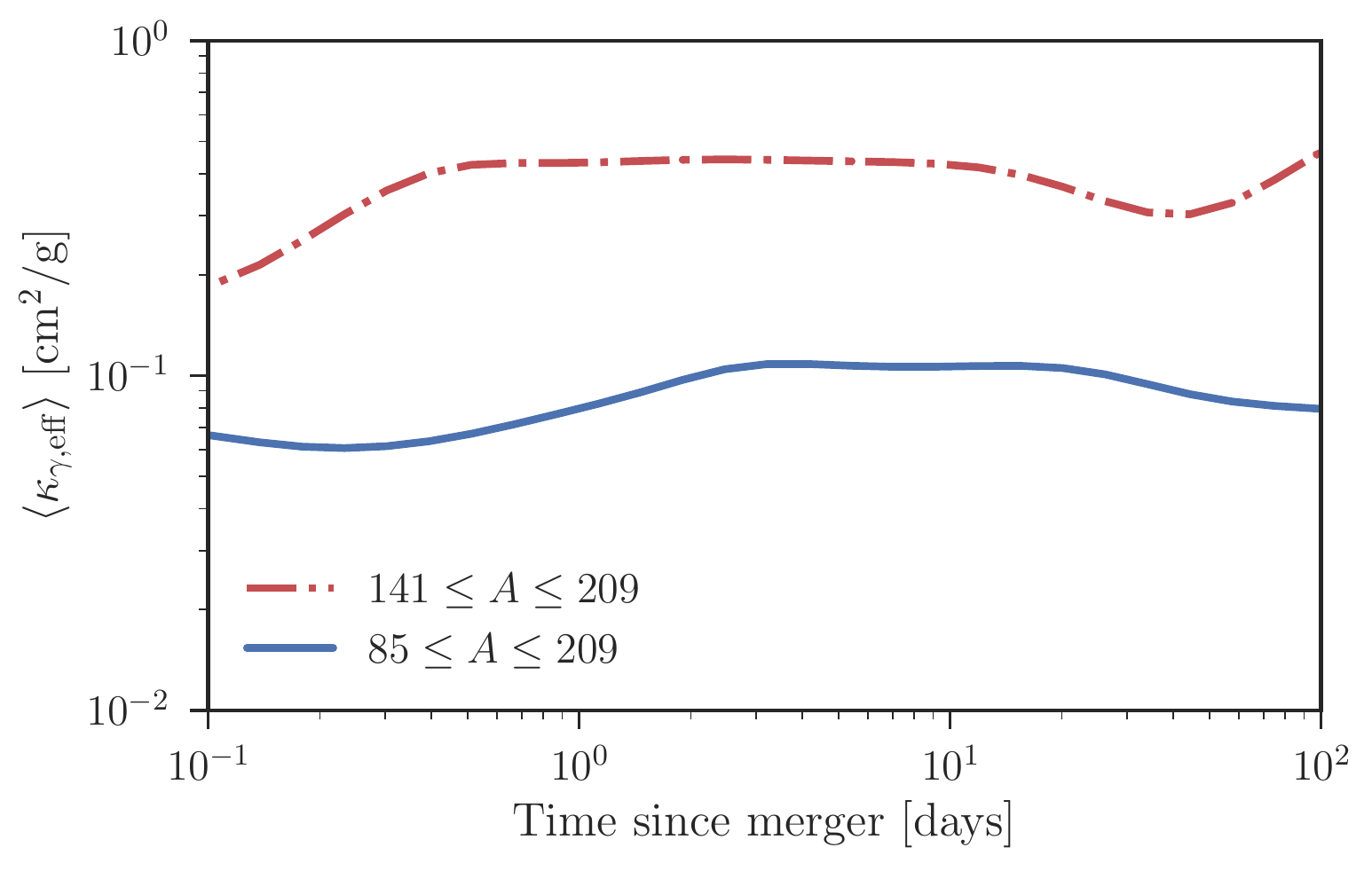}
\end{center}
\caption{
Effective $\gamma$-ray opacity of r-process nuclei as a function time for two different compositions.
\vspace{0.5cm}}
\label{fig:gamma_eff}
\end{figure}

In equation (\ref{fig:stop_gamma}) we approximate $\kappa_{\rm \gamma}(E_i)$ by the geometric mean of the mass attenuation and the energy attenuation opacities. We would like to discuss this approximation shortly. Opacity tables provide two types of opacities, mass attenuation opacity, $\kappa_{\gamma,m}$, which is the opacity for interaction, and energy attenuation opacity, $\kappa_{\gamma,e}$, which is the mass attenuation opacity multiplied by the fraction of its energy that the $\gamma$-ray loses in a single interaction. As evident from figure \ref{fig:stop_gamma}, $\kappa_{\gamma,e}$ is smaller by up to a factor of 2 than $\kappa_{\gamma,m}$. The opacity used in equation (\ref{eq:kappa_gamma}),  $\kappa_{\rm \gamma}$, includes the fraction of energy that is lost during the interactions of the $\gamma$-rays and therefore  $\kappa_{\rm \gamma}$ is lower than the mass attenuation opacity. But, the energy loss takes place also  in multiple interactions and not only in a single interaction and therefore $\kappa_{\rm \gamma}$ is higher than the energy attenuation opacity, and thus, $\kappa_{\gamma,e}<\kappa_{\gamma}<\kappa_{\gamma,m}$. At large optical depth the energy loss is dominated by multiple interactions and therefore $\kappa_{\rm \gamma} \approx \kappa_{\gamma,m}$, while at low optical depth energy deposition is dominated by a single interaction and $\kappa_{\rm \gamma} \approx \kappa_{\gamma,e}$. However, both the high and the low optical depth regimes are of no interest since in the former energy losses are negligible and in the latter they are so severe that electrons dominate the energy deposition by $\beta$-decay and the contribution of $\gamma$-rays can be neglected. In the interesting regime where $\tau_{\rm \gamma,eff} \approx 1$ the value of $\kappa_{\rm \gamma}$ should be calculated numerically, but its value should be somewhere between $\kappa_{\gamma,e}$ and $\kappa_{\gamma,m}$. Given the uncertainty in other parameters, such as the ejecta geometry, there is no much added value in finding the exact value of $\kappa_{\gamma}$ and more so as it depends on the optical depth. Therefore we use the approximation $\kappa_{\gamma} \approx (\kappa_{\gamma,e}\kappa_{\gamma,m})^{1/2}$.  Figure \ref{fig:gamma_eff} shows the time evolution of $\kappa_{\gamma,{\rm eff}}$, as calculated by taking the sum in equation (\ref{eq:kappa_gamma}) over the gamma-rays emitted at any given time by the entire r-process elements (according to the specific composition) and   $\kappa_{\rm \gamma}$ is approximated by the geometric mean of the mass and energy attenuation opacities. Its values around one day are 
$\approx 0.07\,{\rm cm^2/g}$ and $0.4\,{\rm cm^2/g}$ for $85\leq A\leq 209$ and $141\leq A\leq 209$, respectively.

The time scale $t_0$ is estimated as
\begin{eqnarray}
t_{0} & \approx & \left(\kappa_{\rm \gamma,eff}\Sigma_m t^2\right)^{1/2},\\
& \approx & 2.3\,{\rm day}
\left(\frac{C_{\Sigma}}{0.05} \right)^{1/2}
\left(\frac{M_{\rm ej}}{0.05M_{\odot}}\right)^{1/2}
\left(\frac{v_0}{0.1c}\right)^{-1}
\left(\frac{\kappa_{\rm \gamma,eff}}{0.07\,{\rm cm^2g^{-1}}}\right)^{-1/2}.
\end{eqnarray}
Because $25$--$75\%$ of the $\beta$-decay energy (excluding neutrino) goes to $\gamma$-rays, the heat deposition rate decreases by a factor of $\sim 2$ on this time scale.

The fraction of the $\gamma$-ray's energy deposited to the ejecta is then calculated by
\begin{eqnarray}
f_{\gamma}(t) \approx 1-\exp(-\tau_{\rm \gamma,eff})=1-\exp(-(t/t_0)^2).
\end{eqnarray}

\section{Light curve}
Thermal photons created by radioactive heat diffuse out from the ejecta and produce the macronova emission. In this process radiation transfer with the expansion line opacity, which varies with wavelength, is required to calculate the spectrum of macronovae \citep{Kasen2013,barnes2013B,tanaka2013,Wollaeger,Tanaka2019}.   Radiation transfer simulations show that the bolometric light curve can be approximately  obtained by using a grey opacity. For example, the opacity is $\sim 10\,{\rm cm^2/g}$ for lanthanide-rich ejecta \citep{barnes2013B,tanaka2013,Tanaka2017} and $\sim 0.1$-$1\,{\rm cm^2/g}$ for lanthanide-free ejecta \citep{Kasen2015,Tanaka2017,Tanaka2019}. 

Semi-analytic macronova models were developed in the past by various authors \citep[e.g.,][]{li1998ApJ,piran2013MNRAS,Metzger2017LRR,metzger2008ApJ,waxman2018}. These models use various approximations, but they are all built upon the Arnett model for supernova \citep{arnett1982} and they provide good order of magnitude estimates. However, one shortcoming of these models is that they treat properly only radiation that escapes from regions where the diffusion time is shorter than the dynamical time. In the stratified structure of the ejecta contribution of deeper layers, where the diffusion time is longer than the dynamical time, can be non-negligible, especially at early times. Below we present a model, which is also a variant of the Arnett model, that takes proper account of the contribution from these regions.

We model the bolometric light curve and temperature by adding photons diffusing out from  spherical mass shells, where each mass shell is characterized by a mass $m_i$, expansion velocity $v_i$, and grey opacity $\kappa_i$. The internal energy of each mass shell is calculated by solving the first law of thermodynamics of radiation dominated gas:
\begin{eqnarray}
\frac{dE_i}{dt} = -\frac{E_i}{t} + \dot{Q}_i(t) - L_{{\rm rad},i}(t),
\end{eqnarray}
where $E_i$ is the shell's internal energy and $\dot{Q}_i(t)=m_i \dot{Q}_{\rm th}(t)$ is the total heating rate of the shell.  The radiation luminosity of each shell, $L_{{\rm rad},i}$, should account for the following three different regimes: (i) The trapping regime, i.e., $t<t_{{\rm diff},i}$, (ii) the diffusive regime where the radiation escapes over a diffusion time, i.e.,  $vt/c<t_{{\rm diff},i}<t$, (iii) the free streaming regime, i.e., $\tau_i<1$. Here, $\tau_i$ is the optical depth from the $i$'th shell to the observer:
\begin{eqnarray}
\tau_i(t) = \int_{v_it}^\infty \kappa(r)\rho(r)dr,
\end{eqnarray}
and $t_{{\rm diff},i}=\tau_i v_it/c$ is the diffusion time.
In the single zone approximation, the bolometric luminosity can be  approximated as $L\approx E/(t_{\rm diff}+vt/c)$, where $t_{\rm diff}$ is the diffusion time evaluated with the ejecta mass, typical opacity and velocity \citep{Metzger2017LRR}. The sum, $t_{\rm diff}+vt/c$, provides a smooth transition from the trapping regime to the free streaming regime. 
When considering the effect of the ejecta velocity structure, however, the single zone approximation does not work very well because the diffusion time of different regions to the observer is different. In order to take  this effect into account, we calculate the radiation luminosity of each shell as follows.
A good approximation for the energy escape fraction from each shell over one dynamical time when $t_{{\rm diff},i}>t$ is given by 
\citep{Piro2013}
\begin{eqnarray}
f_{{\rm esc},i} & \approx & {\rm erfc}\left(\sqrt{\frac{t_{{\rm diff},i}}{2t}} \right),\label{eq:esc}
\end{eqnarray}
where erfc is the complementary error function. When $t_{{\rm diff},i}<t$, most of the radiation energy escapes, i.e., $f_{{\rm esc},i}\approx 1$, over a diffusion time for $\tau_i\gg 1$ and over the light crossing time for $\tau_i\ll 1$. Since $f_{{\rm esc},i}\approx 1$ for $t_{{\rm diff},i}<t$, equation (\ref{eq:esc}) provides a good approximation at all times and the   escape time of the radiation can be approximated by 
\begin{eqnarray}
t_{{\rm esc},i} & \approx & {\rm min}(t_{{\rm diff},i},t) + \frac{v_it}{c}.
\end{eqnarray}
With these quantities, the luminosity of each shell is approximated as
\begin{eqnarray}
L_{{\rm rad},i} \approx \frac{f_{{\rm esc},i}E_i}{t_{{\rm esc},i}}.
\end{eqnarray}
Then the bolometric luminosity is calculated by adding the contribution of all the shells.

\begin{figure}
\begin{center}
\includegraphics[scale=0.5]{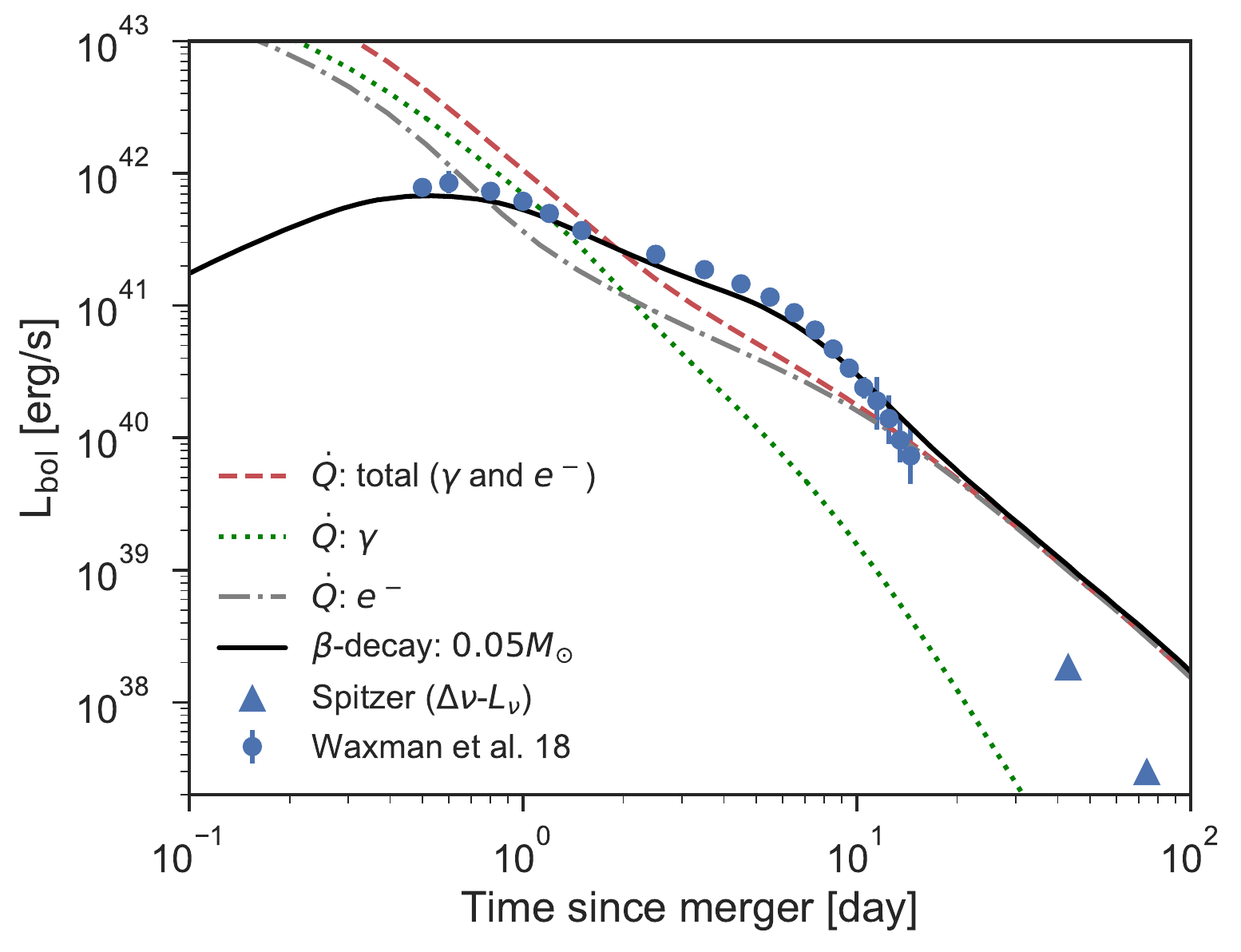}
\includegraphics[scale=0.5]{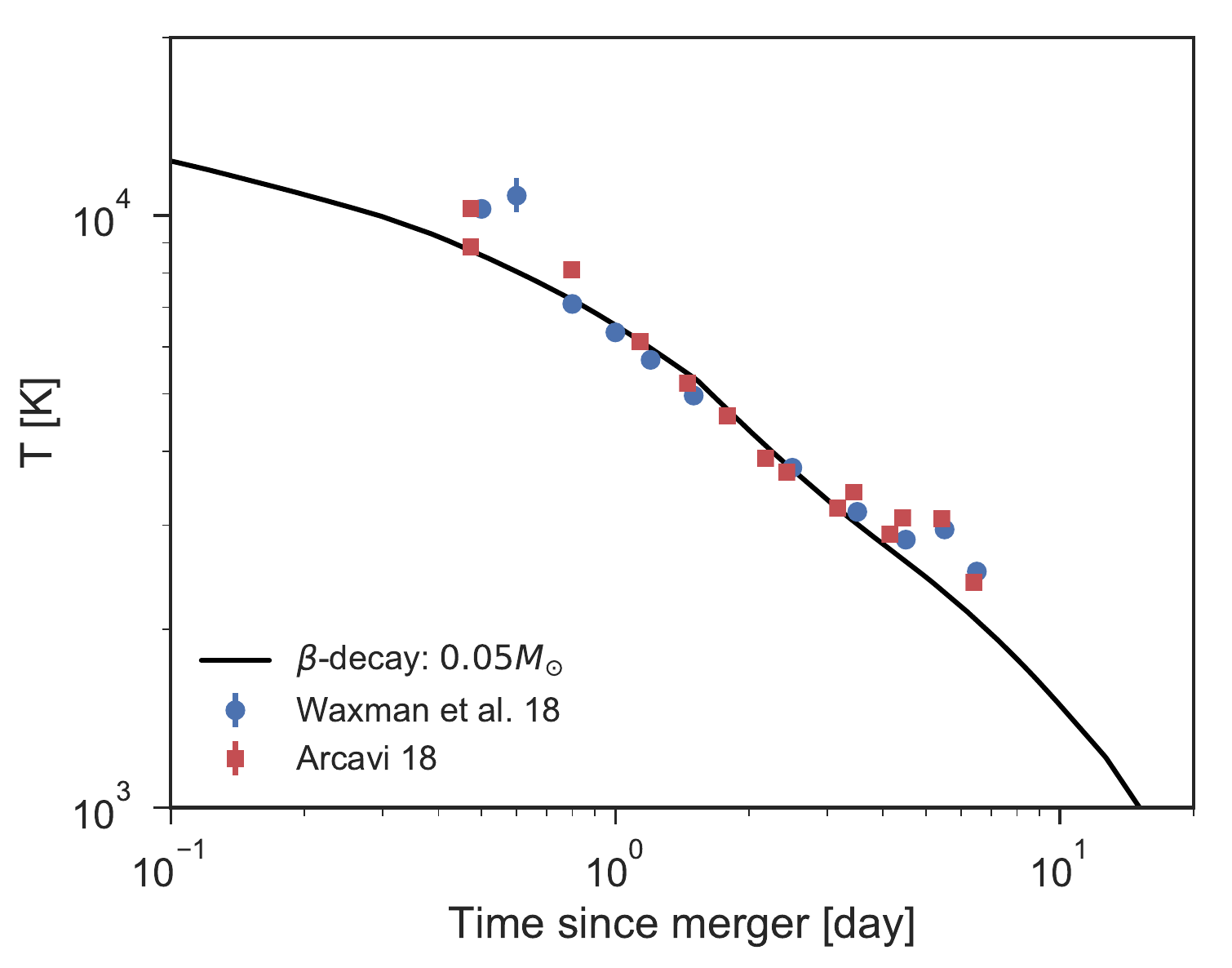}
\end{center}
\caption{
Bolometric light curve and temperature evolution of the macronova associated with GW170817. The total and electron heating rates are also shown. The temperature is evaluated at the photosphere by assuming local thermodynamic equilibrium. Here we use a total ejecta mass of $0.05M_{\odot}$, the beta-decay heating rate with the solar r-process abundance ($85\leq A\leq 209$), and the ejecta profile with $n=4.5$, $v_0=0.1c$ and $v_{\rm max}=0.4c$ (see equation \ref{eq:density}). The opacity is assumed to be $0.5\,{\rm cm^2/g}$ for $v>0.2c$ and $3\,{\rm cm^2/g}$ for $v\leq 0.2c$. The bolometric data are taken from \cite{waxman2018}. The Spitzer $4.5{\rm \mu m}$ detections $\Delta \nu$-$L_{\nu}$ are considered as lower limits on the bolometric luminosity \citep{kasliwal2019} (see discussion in the text). The observed temperature is  shown only up to day 7 \citep{waxman2018,Arcavi2018}, when the spectrum is quasithermal.\vspace{0.5cm}}
\label{fig:Lbol}
\end{figure}

Figure \ref{fig:Lbol} shows the bolometric light curve  of the macronova of GW170817 according to the the analysis of the observation of \cite{waxman2018}. Also shown are the black-body temperature data obtained by \cite{waxman2018} and \cite{Arcavi2018}. The bolometric luminosity shows a roughly steady decay as $\approxprop t^{-1}$ up to day 7 at which point there is a sharp break to a steep decay
as $t^{-3}$. It is important to note that the analysis is robust up to day 7 but at later times it is less certain. The reason is that until day 7 almost the entire emission is within the observable  bands while at later time a significant fraction of the emission is in unobservable IR bands. Moreover, after day 7 also the spectrum is becoming highly non-thermal making any extrapolation of the emission to the IR bands uncertain. Thus, while there is most likely a break around day 7 it is unclear that the post-break slope is as steep as $t^{-3}$. Figure \ref{fig:Lbol} include also two data point which are the IR detection in a single band, $4.5\mu$m, by Spitzer \citep{kasliwal2019}. The spectrum at these times is clearly not thermal (there are simultaneous non-detection at $3.6\mu$m) and cannot be used for a reliable estimate of the bolometric luminosity. Therefore, we consider here only the actual luminosity which was observed within the Spitzer $4.5\mu$m band, which is a strict lower limit of the bolometric luminosity.

Figure \ref{fig:Lbol}  shows also a semi-analytic model of the bolometric light curve and the evolution of temperature at the photosphere. Here, we assume a total ejecta mass of $0.05M_{\odot}$ composed of r-process elements with the solar abundance of $85\leq A \leq 238$. The density profile is assumed to be $\rho \propto v^{-4.5}$ for $0.1c < v < 0.4c$.  To calculate the bolometric light curve, we use radially varying opacity of $0.5\,{\rm cm^2/g}$ for $v> 0.2c$ and $3\,{\rm cm^2/g}$ for $v\leq 0.2c$. 
With these parameters, the calculated light curve and temperature agree with the observed data reasonably well including the early peak at 0.5 day and the break of the light curve around a week\footnote{The black-body temperature at $\sim 0.5$ day depends on how to extrapolate the ultraviolet data at $4$ hours. Including the ultraviolet data reduces the temperature. The two data points at $\sim 0.5$ day from \cite{Arcavi2018}  in figure \ref{fig:Lbol} correspond to with and without the ultraviolet data.}. The reason for this break in our modeling can be understood by comparison of the observed luminosity at any time to the heating rate at the same time. At early times, the photon diffusion wave is  at the outer part of the ejecta so that only a small fraction of the total radioactive deposited energy diffuse out and the emergent luminosity is lower than the total  heating rate. Thus, during this time energy is accumulated within the ejecta and due to adiabatic losses the energy in the ejecta is comparable to the energy deposited over the last dynamical time. On a time scale of a few days, the diffusion wave  proceeds deeper in the ejecta, so the diffusion time through most of the ejecta becomes comparable to the dynamical time. In this phase, all the deposited photons escape to the observer and together with these photons, also radiation that was deposited at earlier times diffuse out from the ejecta, leading to a bolometric luminosity that is higher than the instantaneous heating rate. At later times, where the diffusion wave has crossed all the ejecta, deposited heat escapes on time that is shorter than the dynamical time and the bolometric luminosity  approaches the instantaneous heating rate.  Just before this last phase, there must be a phase where the bolometric light curve  declines faster than the heating rate, corresponding to the break around a week in figure \ref{fig:Lbol}. The same behaviour is seen in all type I SNe where after the peak there is an episode where the bolometric luminosity drops much faster than the $^{56}$Ni heating rate before it convergences to the late time $^{56}$Ni tail. Note that in our model the break is unrelated to any change in the thermalization efficiency. After a week the contribution of the $\gamma$-rays is already negligible while the coupling of the electrons is still efficient. The break in the heating rate that corresponds to inefficient electron coupling is seen only at  $t_{{\rm th},\beta} \approx 30$ days. These results are different than those of \cite{waxman2018,waxman2019} that attribute the break at day 7 to  $t_{{\rm th},\beta}$. The reason for this difference is, at least in part, due to the fact that \cite{waxman2018,waxman2019} assume that the energy of the deposited electrons is $1$\,MeV, while experimental data show that at the relevant time it is typically lower (see figure \ref{fig:beta}), which corresponds to a larger value of $t_{{\rm th},\beta}$.

An interesting point that we find in the attempt to fit the data with different compositions is that including a $\beta$-decay chain, $^{88}$Kr$\rightarrow^{88}$Rb$\rightarrow^{88}$Sr, enhances the peak luminosity, where $^{88}$Kr and $^{88}$Rb have a half-life of $2.83\,{\rm hr}$ and $17.8\,{\rm min}$, respectively. This decay chain releases $\sim 5$ MeV in electrons and $\gamma$-rays. For example, the peak luminosity with $A_{\rm min}=85$ is higher by a factor of $\sim 2$ than that with  $A_{\rm min}=90$. The high peak luminosity of the macronova GW170817 may indicate that this decay chain significantly contributes to the heat around the peak. 

\begin{figure}
\begin{center}
\includegraphics[scale=0.5]{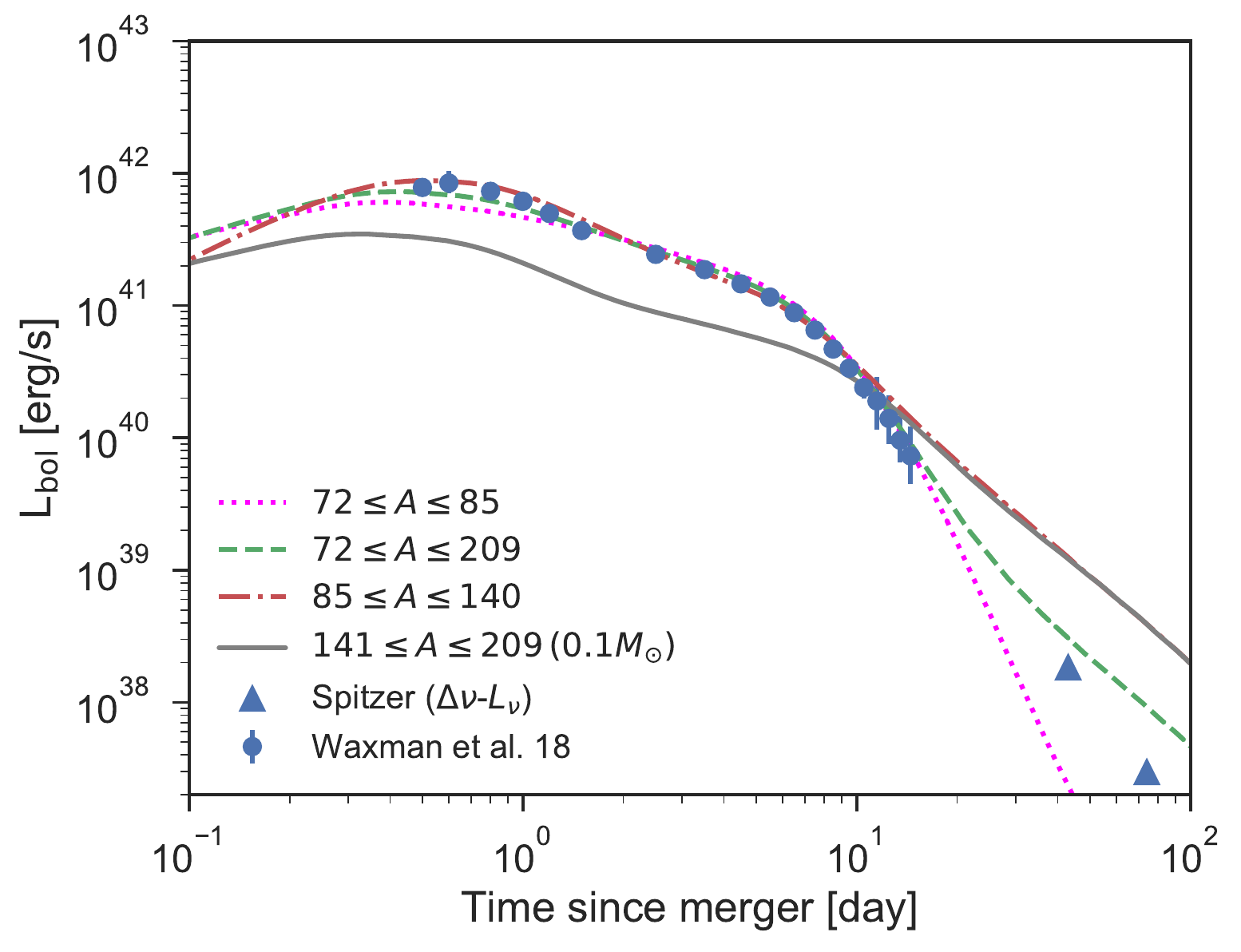}
\end{center}
\caption{
Bolometric light curves for different nuclear compositions.
The ejecta mass is chosen to be $0.1M_{\odot}$ for $141\leq A\leq 209$ and $0.05M_{\odot}$ for the others. The values of the opacity are the followings: $0.3\,{\rm cm^{2}/g}$ ($v>0.18c$) and  $3\,{\rm cm^{2}/g}$ ($v\leq 0.18c$) for $A_{\rm min}=72$, $0.5\,{\rm cm^{2}/g}$ ($v>0.2c$) and  $3\,{\rm cm^{2}/g}$ ($v\leq 0.2c$) for $A_{\rm min}=85$, and
$0.1\,{\rm cm^{2}/g}$ ($v>0.18c$)
$3\,{\rm cm^{2}/g}$ ($v\leq 0.18c$) for $A_{\rm min}=141$. \vspace{0.5cm}
}
\label{fig:Lbol3}
\end{figure}

The dependence of the heating rate on the composition may provide some clues about the ejecta. Figure \ref{fig:Lbol3} shows the bolometric light curves powered by $\beta$-decay with different atomic mass ranges (assuming a solar abundance ratio). The light curve model  with $85\leq A \leq 140$, where there are no elements beyond the second peak, is similar to the one with  $85\leq A \leq 209$. The reason is that the contribution of elements with $A>140$ to the heating is minor. Thus, at least for heating, these elements are not required, although the late time spectrum and color evolution of the macronova GW170817  suggest that the ejecta contains elements beyond the second peak \citep[e.g.,][]{Chornock2017}.
In the case that only the first peak elements are included ($72\leq A\leq85$), the luminosity is too low to reproduce the late-time Spitzer data (see \citealt{kasliwal2019} for details). The reason is that during the first week the heat deposition is dominated by a single chain of the elements with $A=72$ and there are no element with a significant contribution at late times. When heavier elements are added, $72\leq A \leq 209$, the emission at late time is brighter and marginally consistent with the strict Spitzer lower limits. The reason for the rather low late-time heating  (compared to the case with $85\leq A \leq 209$) is that also here the large mass carried by first peak elements that do not contribute to the late time emission is coming on the expense of the heavier elements that contribute to the late-time heating. Given that the Spitzer lower limits account only for the emission seen within the Spitzer band, it is most likely that the actual bolometric luminosity is at least a factor of a few brighter than these lower limits and therefore it is most likely that the ejecta did not contain a significant fraction of first peak elements. Finally, when only elements beyond the second peak are included ($140\leq A\leq 209$), the luminosity at early times is lower by a factor of $\sim 5$ than the observed data. This suggests that while elements above the second peak are probably present in the ejecta (based on their opacity signature), the total ejecta mass is dominated by elements with atomic mass $85\leq A\leq140$.

\begin{figure}
\begin{center}
\includegraphics[scale=0.5]{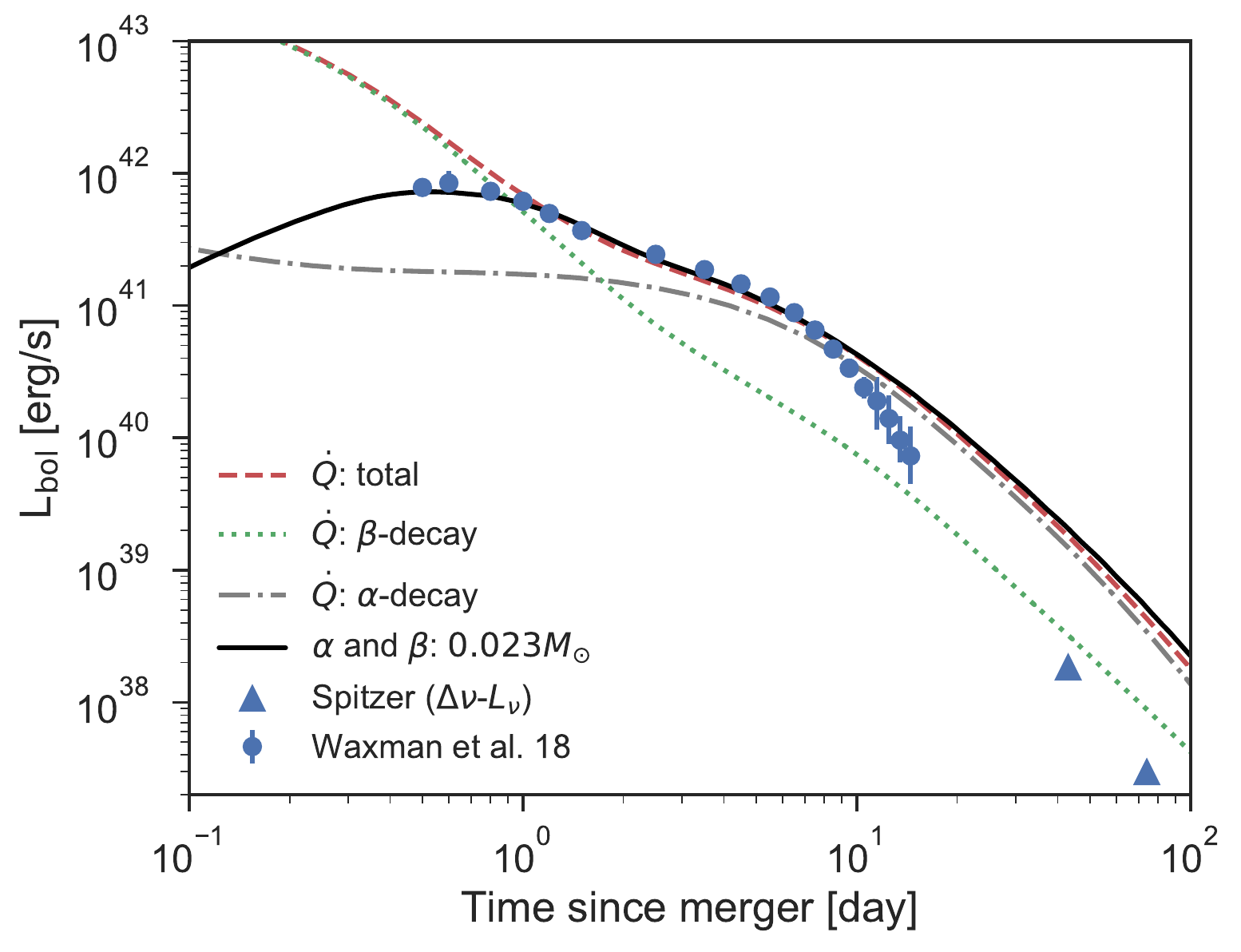}
\includegraphics[scale=0.5]{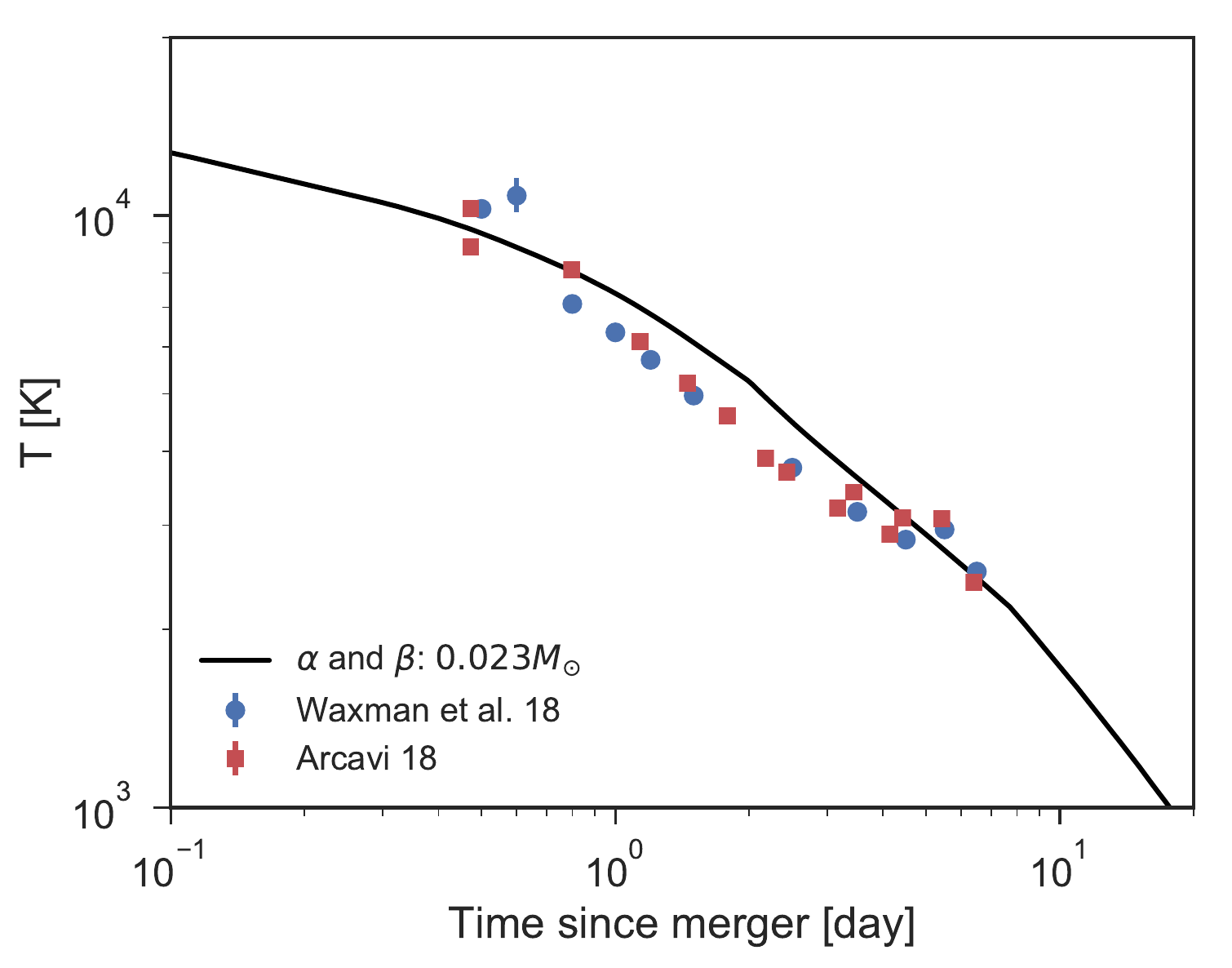}
\end{center}
\caption{
Same as Fig. \ref{fig:Lbol} including $\alpha$-decay heating and a total ejecta mass of $0.023M_{\odot}$. The initial abundance of $(Y(^{222}{\rm Rn}), Y(^{223}{\rm Ra}), Y(^{224}{\rm Ra}), Y(^{225}{\rm Ra}))=(4.0\cdot 10^{-5},2.7\cdot 10^{-5},4.1\cdot 10^{-5},2.7\cdot 10^{-5})$.
}
\label{fig:Lbol2}
\end{figure}

Figure \ref{fig:Lbol2} depicts the bolometric light curve and temperature in the case that $\alpha$-decay heating is included assuming the abundances of $\alpha$-decaying nuclei used for figure \ref{fig:h2}. Because the heating rate at later times is significantly enhanced by the $\alpha$-decay contribution, the total ejecta mass  required to fit the data is reduced to $\approx 0.023M_{\odot}$. Here, we use the density profile same to the above and the opacity of $0.5\,{\rm cm^2/g}$ for $v> 0.14c$ and $3\,{\rm cm^2/g}$ for $v\leq 0.14c$.
In this model, the light curve at $1 \lesssim t\lesssim 10$ days declines with $\propto t^{-1}$
resulting from that the $\alpha$-decay heating kicks in around $2$ days. Then the model light curve turns to declines as $\propto t^{-2.8}$ due to the thermalization inefficiency. However, the observed light curve falls more quickly than the model light curve, although this may be a result of underestimate of the observed bolometric luminosity at $t \gtrsim 7$ days.

\section{Ejecta mass estimate based on the Katz integral}
Estimate of the ejecta mass that uses light curve modeling are degenerated with the opacity, heating rate, density profile, as well as the outflow geometry and the viewing angle. 
\cite{katz2013} suggest a powerful method to obtain the total mass of radioactive elements, $M_{\rm rad}$, from observed bolometric light curve data, $L_{\rm bol}(t)$, as long as the  heat deposition rate is known.  
The following relation between the heating rate and the bolometric light curve should be valid for all times $t\gg t_{\rm f}$:
\begin{eqnarray}
M_{\rm rad}\int_0^{t} t'\cdot \dot{Q}_{\rm th}(t') dt' = \int_0^{t} t'\cdot L_{\rm bol}(t') dt',\label{eq:katz}
\end{eqnarray}
where $t_{f}$ is the time where the diffusion wave crosses the entire ejecta and the bolometric luminosity approches the instantanouos heating rate, i.e., $L_{bol}(t \gg t_{f}) \approx \dot{Q}_{\rm th}(t)$. Since the heating rate $\dot{Q}_{\rm th}$ depends  on the ejecta composition,  $M_{\rm rad}$ is determined for a given ejecta composition (see \citealt{nakar2016} for an application to core-collapse supernovae). We emphasize  that this method is fully independent of the opacity, which is the most uncertain quantity, and it is also almost independent of the ejecta geometry and velocity profile\footnote{The only dependence on the ejecta structure is through the thermalization time, which affects the heat deposition rate
on the left-hand side of equation (\ref{eq:katz}).}.  
The light curve of the  macronovae of GW170817 decline rapidly ay $t>7$ day and at the same time the spectrum becomes non-thermal suggesting that by that time the diffusion wave crossed most of the ejecta and therefore the available bolometric light curve data is sufficient in order to use this method to estimate $M_{\rm rad}$  and under the assumption that the ejecta is composed entirely by r-process elements $M_{\rm ej}=M_{\rm rad}$.

\begin{figure}
\begin{center}
\includegraphics[scale=0.55]{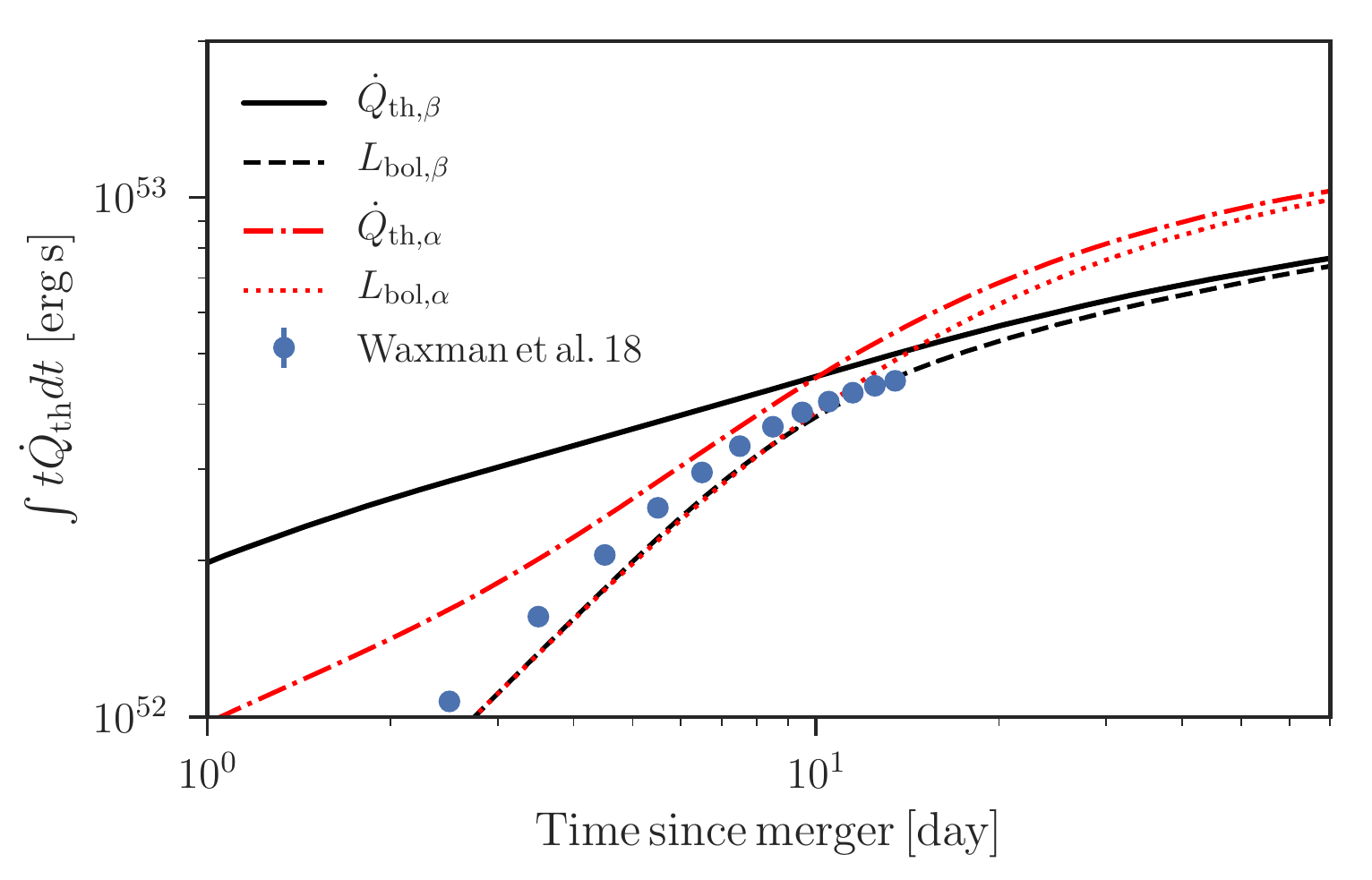}
\includegraphics[scale=0.55]{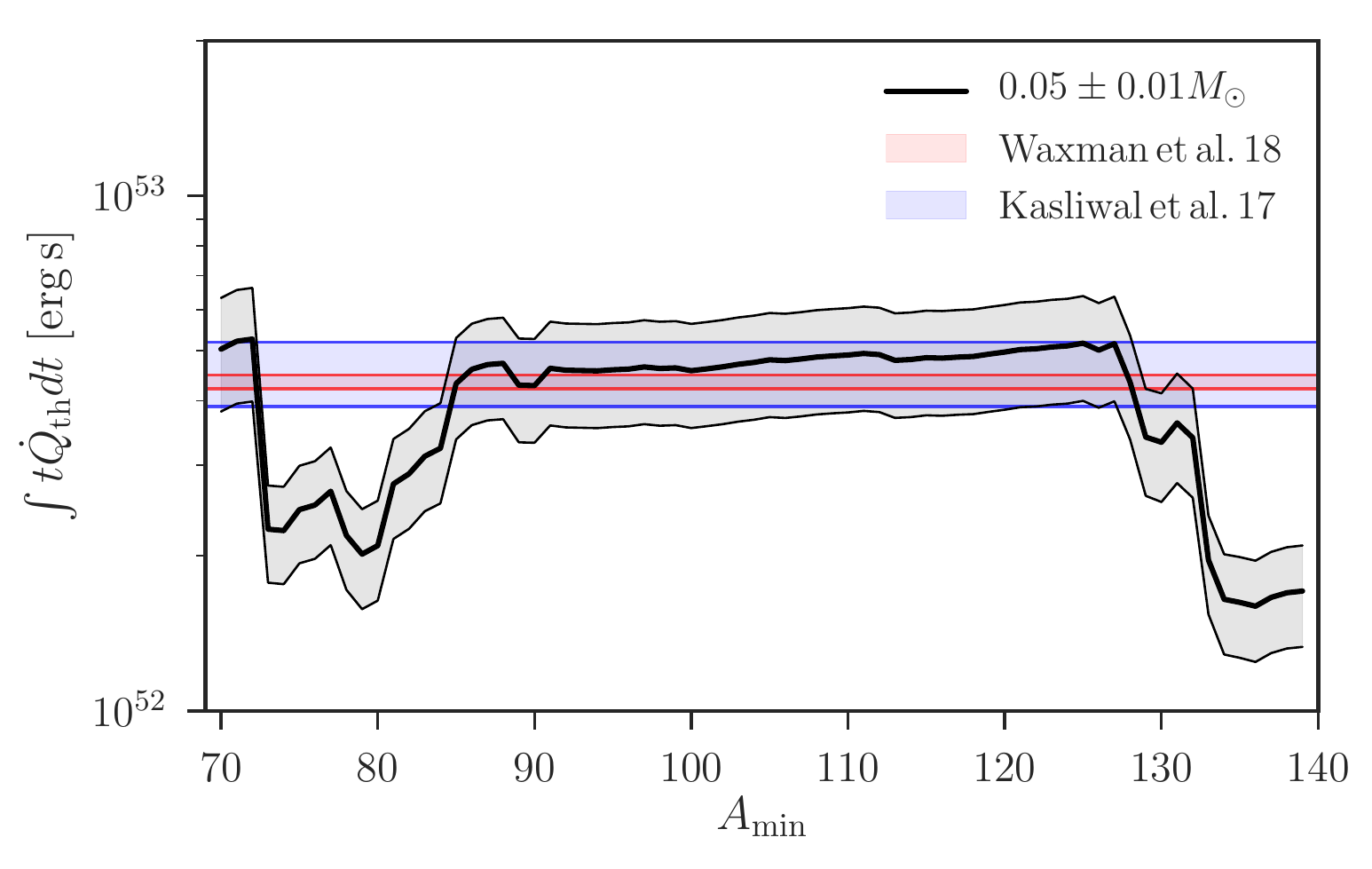}
\end{center}
\caption{
Katz integral of heating rates and light curves ({\it left}) and Katz integral up to $12.5$ days as a function of the minimum atomic mass number ({\it right}). The solar r-process abundance with $A\geq A_{\rm min}$ is used. The total r-process mass is set to be $0.05\pm 0.01M_{\odot}$ and
the ejecta velocity profile with $v_0=0.1c$, $v_{\rm max}=0.4c$ and $n=4.5$ is used. 
Here we consider only $\beta$-decays as a heat source.
Also shown as horizontal bars are the time-weighted integral of the observed bolometric light curve of the macronova GW170817 (the right hand side of equation \ref{eq:katz}). The bolometric data are taken from \cite[SED integration]{waxman2018} and \cite{Kasliwal2017}.\vspace{0.5cm}
}
\label{fig:katz}
\end{figure}

Figure \ref{fig:katz} ({\it left}) depicts the time-weighted integral of the heating rate and bolometric luminosity. Here, we use the $\beta$-decay and $\alpha$- and $\beta$-decay models shown in figure \ref{fig:Lbol} and \ref{fig:Lbol2}. The integral of the heating rate approaches that of bolometric luminosity around $10$ days for the $\beta$-decay model and $5$ days for the $\alpha$- and $\beta$-decay model. 
Figure \ref{fig:katz} ({\it right}) shows the Katz integral as a function of the minimum atomic mass number, where we assume the solar r-process abundance pattern with $A_{\rm min}\leq A \leq 209$. The grey region shows the integral with the ejecta mass of $0.05\pm 0.01M_{\odot}$ at $12.5$ days. Also shown as red and blue horizontal bars are the right hand side of equation (\ref{fig:katz}) based on the observed bolometric data of the macronova in GW170817 taken from \cite{waxman2018} and \cite{Kasliwal2017}, respectively. The comparison between these two quantities suggests that the ejecta mass in this event is $\approx 0.05M_{\odot}$ for $A_{\rm min}\leq 72$ and $85\leq A_{\rm min}\leq 130$ with the solar r-process abundance pattern.

\section{Discussion}
\subsection{Abundance pattern of light r-process elements}
Here we consider the elemental abundance pattern of merger ejecta based on a hypothesis that r-process elements of  r-process enhanced metal poor stars are predominantly produced by neutron star mergers.
The observations of these  stars reveal that their elemental abundance patterns beyond the second r-process peak ($A\gtrsim 140$)  are practically  indistinguishable among them and in agreement with the solar pattern. 
The abundances of the elements between the first and second peaks ($90\lesssim A\lesssim 120$) relative to the europium abundance are scattered around the solar pattern within $1\,{\rm dex}$. Notably, r-process enhanced metal poor stars contain systematically less amounts of Gd and Ge (the first r-process peak) by  $\gtrsim 1$ dex  than that expected from the solar abundance pattern. If the hypothesis is correct, these observational facts suggest that mergers produce r-process elements with an abundance similar to the solar pattern with a minimum atomic mass number of $75 \lesssim  A_{\rm min}\lesssim 85$.

R-process nucleosynthesis calculations show a clear dependence of the abundance pattern on the initial electron fraction, $Y_e$ \citep{korobkin2012MNRAS,perego2014MNRAS,wanajo2014ApJ,lippuner2015ApJ,wanajo2018}. Heavy elements beyond the second peak are produced for $Y_e\lesssim 0.2$.
When $Y_e\sim 0.25$, synthesized nuclei are mostly those around the second peak. The first peak elements are predominantly produced for $Y_e\gtrsim 0.35$.
Numerical simulation of dynamical ejecta \citep{freiburghaus1999ApJ,bauswein2013ApJ,sekiguchi2015PRD,Foucart2016,radice2016MNRAS,Radice2018,Bovard2017} and disk wind \citep{fernandez2013MNRAS,fernandez2015MNRAS,metzger2014MNRAS,perego2014MNRAS,just2015MNRAS,fujibayashi2017,Siegel2018}
show that $Y_e$ is broadly distributed in a certain atomic number range and the abundance patterns are typically consistent with the solar pattern with a minimum atomic mass of $A_{\rm min}\sim 80$--$120$. Note, however, that lighter elements are synthesized more when neutrino absorption significantly changes the electron fraction. This occurs when the mass of the accretion disk is sufficiently large \citep{just2015MNRAS,Miller2019} and a long-lived massive neutron star is formed \citep{metzger2014MNRAS,Lippuner2017, Shibata2017}. Our results, which are based on the heating rate that is needed to explain the observed light curve, suggest that if the composition of the ejecta from GW170817 resembles the solar abundance pattern then  $A_{\rm min} \geq 85$ and its mass was dominated by elements in the range  $85<A <140$. This result is consistent with the estimates of the opacity, that suggest the a significant fraction of the ejecta was lanthanides poor and that in the part of the ejecta that contained lanthanides the fraction was comparable or lower than the solar abundance \citep[e.g.][]{Kasen2017,Tanaka2017,waxman2018}.

\subsection{Abundance pattern of super-heavy r-process elements}
The macronova heating rate is potentially dominated by $\alpha$-decay at later times
depending on the initial abundance of super-heavy nuclei of $222\leq A\leq 225$ \citep{wu2019}. For instance, the $\alpha$-decay heating rate is more powerful than
$\beta$-decay by a factor of $\sim 5$ after $10$ days when $Y(^{A}{\rm X})/Y({\rm Eu})\sim 1$, where $^{A}{\rm X}$ is $^{222}$Rn, $^{223}$Ra, $^{224}$Ra, and  $^{225}$Ra. 
However, the contribution of  $\alpha$-decay heating is still under debate since different nuclear mass models predict different initial abundances of these nuclei \citep{wu2019,wanajo2018}.
Here we briefly discuss the initial abundance of $\alpha$-decaying nuclei  inferred from the measurements of Pb of r-rich stars.  The measured abundance ratio of Pb to Eu of r-rich stars,  $Y({\rm Pb})/Y({\rm Eu})$, is $\approx 4$ for typical r-rich stars and $\approx 5$ for the solar r-process pattern. The abundance of
Pb  created via $\alpha$-decays should not exceed this value. If all Pb results from the decay chains starting with $^{222}$Rn, $^{223}$Ra, and $^{224}$Ra, their initial abundances are $Y(^{A}{\rm X})/Y({\rm Eu})\sim 1$ and this case corresponds to the upper limit on the $\alpha$-decay contribution:
\begin{eqnarray}
\dot{Q}_{{\rm th,}\alpha}(t) \lesssim 5\dot{Q}_{{\rm th,}\beta}(t)~~~~~~~~(t\gtrsim 10\,{\rm days}).
\end{eqnarray}
In reality, however, it is more likely that Pb is created by a larger number of decay chains. 
If we  consider that $\alpha$-decaying elements are synthesized with a flat abundance pattern extending up to $A=250$ as a zeroth-order estimate of the production ratios (see, e.g., \citealt{eichler2015ApJ} for the abundances of heavy nuclei after r-process),
 we obtain $Y(^{A}{\rm X})/Y({\rm Eu})\sim 0.12$ corresponding to that the fraction of the heating rate of $\alpha$-decay to that of $\beta$-decay is about $0.6$.  


\section{Summary}
We study the radioactive power of r-process nuclei and  thermalization of $\gamma$-rays, electrons, $\alpha$-particles, and fission fragments. We calculate the coupling of the gamma-rays and the charged particles to the ejecta material using  experimental data of the injection energies and lifetimes. We find that the optical depth for gamma-rays depends on the ejecta composition and on the time after the merger and that it is typically higher than the one found for $^{56}$Ni in ejecta of type Ia SN. About 1 day after the merger its average value is $\kappa_{{\rm \gamma,eff}} \approx 0.07 {\rm ~cm^2/gr}$ for light r-process elements ($A \lesssim 140$) and $\kappa_{{\rm \gamma,eff}} \approx 0.4 {\rm ~cm^2/gr}$ for heavy r-process elements ($A \gtrsim 140$).
For ejecta with $\sim 0.05M_{\odot}$  of mostly light r-process elements and a minimum velocity of about $0.1$c, as inferred for the ejecta of GW170817, the gamma-rays decouple from the ejecta after about 2 days.
We calculate the time at which the coupling of charged particles becomes inefficient. For $\beta$ and $\alpha$ particles this time depends on the injection energies as $E_{i,0}^{-0.5}$, and since the injection energies of $\beta$ particles vary by more than an order of magnitude between different elements, this dependence should be taken into account. For an ejecta mass of $0.05M_{\odot}$ and minimum velocity of  $0.1c$, we find that the $\beta$-decay heating rate starts to deviate from the radioactive power at $\sim 10$ day and gradually approaches $\propto t^{-2.8}$ after the thermalization time $t_{\rm th,\beta}\sim 55$ day. We find that the asymptotic heating decay rate after the coupling becomes inefficient is  $\propto t^{-2.8}$ for $\alpha$ and $\beta$ particles  (in agreement \citealt{waxman2019}) and $t^{-3}$ for  spontaneous fission ($^{254}$Cf). Note that for $\alpha$ and fission particles these scalings are valid on time scales not much longer than the mean life-time of their parent nuclei.

We wrote a code that computes the heating rate for a given initial nuclear abundance and outflow parameters. This code takes into account the specific gamma-ray and electron injection energies of each element and follow their instantaneous heating rate numerically. The code is publicly available at 
\url{https://github.com/hotokezaka/HeatingRate}.

We also present an analytic modeling to calculate macronova light curves arising from a homologously expanding ejecta with a velocity gradient. Our model accounts for the photon diffusion from different mass shells with different expansion velocities.
With the $\beta$-decay heating rate of r-process nuclei, we demonstrate that  an ejecta model composed of low opacity material, $\kappa\approx 0.5\,{\rm cm^2/g}$, in the outer part and higher opacity material, $\kappa\approx 3\,{\rm cm^2/g}$, in the inner part reproduces well
the macronova observations of GW170817.  We interpret a break in the observed bolometric light curve around a week as a result of the photon diffusion wave crosses a significant fraction of the ejecta rather than a result of a thermalization break as suggested by \cite{waxman2018}. 
This light curve modeling is also included in the code. 

We use the comparison between a time-weighted integral of the heating rate  and that of the observed bolometric light curve to estimate
the total mass of r-process nuclei produced in GW170817.  This method allows us to obtain an estimate of the ejecta mass that is completely independent of the uncertain ejecta opacity and that is only weakly dependent on the outflow geometry. Assuming that the macronova in GW170817 is powered by $\beta$-decay of r-process elements with solar abundance pattern over some range of atomic masses,
we obtain a total mass of $\approx 0.05M_{\odot}$ for a minimum atomic mass number of $A_{\rm min}\leq 72$ and $86\leq A_{\rm min} \leq 130$.
In the case that $A_{\rm min}$ is outside of the above ranges, a larger ejecta mass is required to explain the data. When the fit to the entire light curve is also considered we find that the late time heating rate suggests that the ejecta did not contain a significant fraction of first peak elements, (i.e.,  $A_{\rm min} \gtrsim 85$). Instead it was probably dominated by elements at and below the second peak, i.e., $85 \leq A \leq 140$. The ejecta probably contained elements above the second peak (as suggested by the IR opacity at late times), but these elements did not have to play a major role in the heating of the ejecta.


\section*{Acknowledgements}
We thank Shinya Wanajo, Eli Waxman, and Meng-Ru Wu for useful discussions. K. H. is supported  by Lyman Spitzer Jr. Fellowship at Department of Astrophysical Sciences, Princeton University. E.N. was partially supported by the Israel Science Foundation (grant 1114/17) and by an ERC consolidator grant (JetNS).

\appendix
\section{Stopping power of charged particles}
Here we describe the energy loss of fast particles due to ionization and excitation and Coulomb collision with thermal electrons.
The stopping power for electrons due to ionization and excitation is given by the Bethe formula:
\begin{eqnarray}
K_{{\rm st},e} = \frac{4\pi Z_2e^4}{Mm_e v^2} \left[\ln\left(\frac{\gamma\sqrt{m_ev^2E}}{2^{1/2}\langle I\rangle} \right)-
\left(\frac{1}{\gamma}- \frac{1}{2\gamma^2}\right)\ln2 +\frac{1}{2\gamma^2} +\frac{1}{16}\left(1-\frac{1}{\gamma}\right)^2\right],\label{eq:ele}
\end{eqnarray}
where  $E$ is the electron's kinetic energy, $Z_2$, $M$, and $\langle I \rangle$ are the atomic number, the atomic mass, and the mean excitation energy  of the stopping medium.

The stopping power for fission fragments due to ionization and excitation at low energies is approximated by \citep{Mukherji1974}
\begin{eqnarray}
K_{\rm st,sf} \approx 8.4\cdot 10^{3}\frac{f(Z_2)}{A_2}f(Z_1)^{3/2} \left(\frac{v}{10^8{\rm cm/s}}\right) \,{\rm MeV cm^2/g},
\end{eqnarray}
where $Z_1$ is the atomic number of a fission fragment, $A_2$ is the mass number of the stopping medium, and
\begin{eqnarray}
f(Z) \approx 0.28Z^{2/3}.
\end{eqnarray}

The stopping power for a charged particle moving at $v$ in a thermal plasma with thermal velocity $v_{\rm th}\ll v < c/137$ is given by \citep{bohr}
\begin{eqnarray}
K_{\rm st,th} \approx \frac{4\pi z^2e^2 \chi }{M m_e v^2}\ln\left(\frac{1.123m_ev^3}{ze^2\omega_p} \right),
\end{eqnarray}
where $\omega_p$ is the plasma frequency, 
$\chi$ is the free electron fraction. Finally, we use the ASTAR database (\url{http://physics.nist.gov/Star}) for the stopping power for $\alpha$-particles due to ionization and excitation.

\section{Energy loss of charged particles in ejecta}
Equation (\ref{eq:cool}) can be solved analytically \citep{kasen2018,waxman2019}. In the following, we describe the solutions used to calculate thermalization in this work.  Introducing dimensionless variables, $\epsilon=E/E_0$ and
$\tau = t/t_{\rm th}$ , where $E_0$ is the initial energy  and $t_{\rm th}^2=C_{\rho}c(\kappa_{\rm st}\beta)_0M_{\rm ej}/v_{\rm ej}^3E_0$, equation (\ref{eq:cool}) is rewritten as
\begin{eqnarray}
\frac{d\epsilon}{d\tau}=-\frac{\epsilon^{-a}}{\tau^3} -x_{\rm ad}\frac{\epsilon}{\tau},
\end{eqnarray}
where the stopping power at $E_0$ is $(K_{\rm st}\beta)_0$  and $K_{\rm st}\beta=(K_{\rm st}\beta)_0\epsilon^{-a}$.  If we neglect the energy dependence of $x_{\rm ad}$, the energy of a fast particle injected at $\tau_0$   is found as ($a\neq 1$)
\begin{eqnarray}
\epsilon(\tau;\tau_0) & = & \left(\frac{\tau}{\tau_0}\right)^{-x_{\rm ad}}
\left[1 - \frac{(1+a)\tau_0^{-2}}{x_{\rm ad}(1+a)-2}\left(\left(\frac{\tau}{\tau_0}\right)^{x_{\rm ad}(1+a)-2}-1\right) \right]^{1/(1+a)}.
\end{eqnarray}
The first part corresponds to the adiabatic energy loss and the second part corresponds to the collisional energy loss. 
For $a=1$ (fission fragments), we have
\begin{eqnarray}
\epsilon(\tau;\tau_0) = \left(\frac{\tau}{\tau_0}\right)^{-x_{\rm ad}}
\exp\left[-\frac{1}{2\tau_0^2}\left(1-\frac{\tau_0^2}{\tau^2} \right) \right].
\end{eqnarray}

\end{document}